\documentclass[journal,12pt,draftclsnofoot,onecolumn]{IEEEtran}
\usepackage{cite}

\ifCLASSINFOpdf
  \usepackage[pdftex]{graphicx}
\else
  \usepackage[dvips]{graphicx}
  \usepackage[export]{adjustbox}
\fi
\usepackage[cmex10]{amsmath}
\usepackage{mathabx}
  \usepackage[caption=false,font=footnotesize]{subfig}
\usepackage{float}

\begin{document}
%
\title{Extended OOBE Comparisons for OFDM, GFDM and WCP-COQAM at Equal Spectral Efficiency}
%
%
%

        

\author{Ali~Bulut~\"{U}\c{c}\"{u}nc\"{u},~\IEEEmembership{Student Member,~IEEE},
        and Ali~\"{O}zg\"{u}r~Y{\i}lmaz,~\IEEEmembership{Member,~IEEE}
\thanks{The authors are with the Department of Electrical and Electronics Engineering, Middle East Technical University, Ankara, Turkey (e-mail: ucuncu@metu.edu.tr, aoyilmaz@metu.edu.tr)}}

%
%

\markboth{}%
{}
%



\maketitle

\begin{abstract}
Generalized frequency division multiplexing (GFDM), windowed cyclic prefix circular offset quadrature amplitude modulation (WCP-COQAM) and orthogonal frequency division multiplexing (OFDM) are among the candidate 5G modulation formats. In this study, we present additional results for the OOBE comparisons between OFDM, GFDM and WCP-COQAM under equal or unequal spectral efficiency conditions for various simulation scenarios. \end{abstract}

\begin{IEEEkeywords}
GFDM, out-of-band emission (OOBE), carrier frequency offset (CFO), WCP-COQAM, OFDM.
\end{IEEEkeywords}

%
\IEEEpeerreviewmaketitle

\section{Introduction}
\IEEEPARstart{O}{rthogonal} frequency division multiplexing (OFDM) is used in recent communication standards owing to  its simple implementation and robustness in inter-symbol interference (ISI) channels\cite{Schulze_OFDM}. However, it has high levels of out-of-band emissions (OOBE) owing to its rectangular pulse shape in time domain. Moreover, when orthogonality between subcarriers is destroyed, it can cause high inter-carrrier-interference (ICI). These disadvantages of OFDM results in the proposal of alternative modulation schemes for 5G and other future systems\cite{farhang2011ofdm}, \cite{BER_sens}. Among the alternatives, there are generalized frequency division multiplexing (GFDM) \cite{GFDM_5G} and windowed cyclic prefix circular offset quadrature amplitude modulation (WCP-COQAM) \cite{lin2014multi}. In this work, we present OOBE results for OFDM, GFDM and WCP-COQAM under equal or unequal spectral efficiency conditions for a wide range of simulation scenarios. Moreover, we also present the error rate performances under CFO for the three modulation types.

\section{Simulations}
The base simulation scenario parameters are specified in Table \ref{table:sim_param}.

\begin{table}[H]
\renewcommand{\arraystretch}{1.2}%
\caption{Simulation Parameters.}
\label{table:sim_param}
\begin{center}
\begin{tabular}{  l | c  }
\hline
Parameter  & Value\\
\hline
Total number of subcarriers (K) & 128 (Part A) or 1152 (Part B)    \\
No. of guard subcarriers  & 52 or 468 \\
Number of time slots (M)     & 9     \\
Pulse Shape     	      & RC (with roll-off=0.1)    \\
Constellation             & 4-QAM\\
CP length				  & 32     \\
Windowing				  & Hanning, 18 samples from both sides \\
Spectral estimation method& Periodogram \\
Interpolation filter type & RC pulse with roll of 0.1 \\
Interpolation filter duration & 81 symbols \\
No. of Monte-Carlo simulations& 900 \\
Sampling Rate & 1 MHz \\
\hline
\end{tabular}
\end{center}
\end{table}

\section{Variation of the base scenario parameters}
In the simulations presented in this document, some of the parameters in Table~\ref{table:sim_param} is changed while all remaining parameters are maintained to be the same. Only the parameters that are changed from the values specified in Table~\ref{table:sim_param} will be mentioned. Moreover, non-contiguous spectrum case will also be examined.

PSD estimates of the three modulation schemes (OFDM, GFDM and WCP-COQAM) are found by means of periodogram, as the average of the PSDs calculated by taking the DFT of each interpolated (by 6) OFDM, GFDM or WCP-COQAM symbol as performed in \cite{banelli_OFDM_nonlin}. A significant point may be about how interpolation is performed to increase the sampling rate to be able to see the out-of-band (OOB) portion of PSDs. In our base scenario simulations, time-domain signals are sixfold oversampled using an RC filter with a length of 81 symbols.
However, after filtering, the samples are truncated from both sides (from the beginning and end) of OFDM, GFDM or WCP-COQAM symbols, in order that the total number of samples is 6 times the
number of samples in the signals that are not oversampled. Otherwise, depending on the length of the interpolation filter, OFDM, GFDM or WCP-COQAM symbols would leak to the
neighbouring symbols, which will require additional cyclic prefix length that will decrease the overall spectral efficiency. Furthermore, if such a truncation is not made, interpolation filter itself would give the effect of windowing, which can convey results that windowing has little or no effect in terms of OOB emissions.

Regarding the CP length, since the maximum delay spread of the channel model used (static inter-symbol interference (ISI) COST-207 hilly terrain channel model) is $20 \mu s$, which corresponds to 20 samples (note that the sampling rate is taken to be 1MHz), single-tap zero-forcing type equalization will be possible owing to the fact that the delay spread is less than the cyclic prefix length. When windowing is applied using the approach described in \cite{Schulze_OFDM}, the CP length will be even longer than 32.

In the following sections, OOB emissions under equal and unequal spectral efficiency conditions and error rate performances under carrier frequency offset (CFO) with static ISI COST-207 hilly terrain channel model are presented.

\pagebreak
\subsection{\textbf{Variations in the Pulse Shape Type}}
In this section, RC type pulse shape that is used for GFDM and WCP-COQAM in the base simulation scenario is changed with different pulse-shape types. Corresponding OOB emissions and error rate performances under CFO are plotted. $\Delta f$ on the symbol error rate (SER) performance plots corresponds to subcarrier spacing and $F_s$ is the sampling rate before upsampling by 6 is performed.
\subsubsection{\textbf{RRC pulse}}~\\ \hspace{30pt}\underline{OOB Emissions:}

\begin{figure}[H]
\centering
\begin{minipage}[b]{0.42\linewidth}
\graphicspath{ {C:/Bulut/Yuksek_Lisans/Yayinlar/OOB_GFDM/OOB_plots/OOB_plots_v2/} }
\includegraphics[width=\columnwidth]{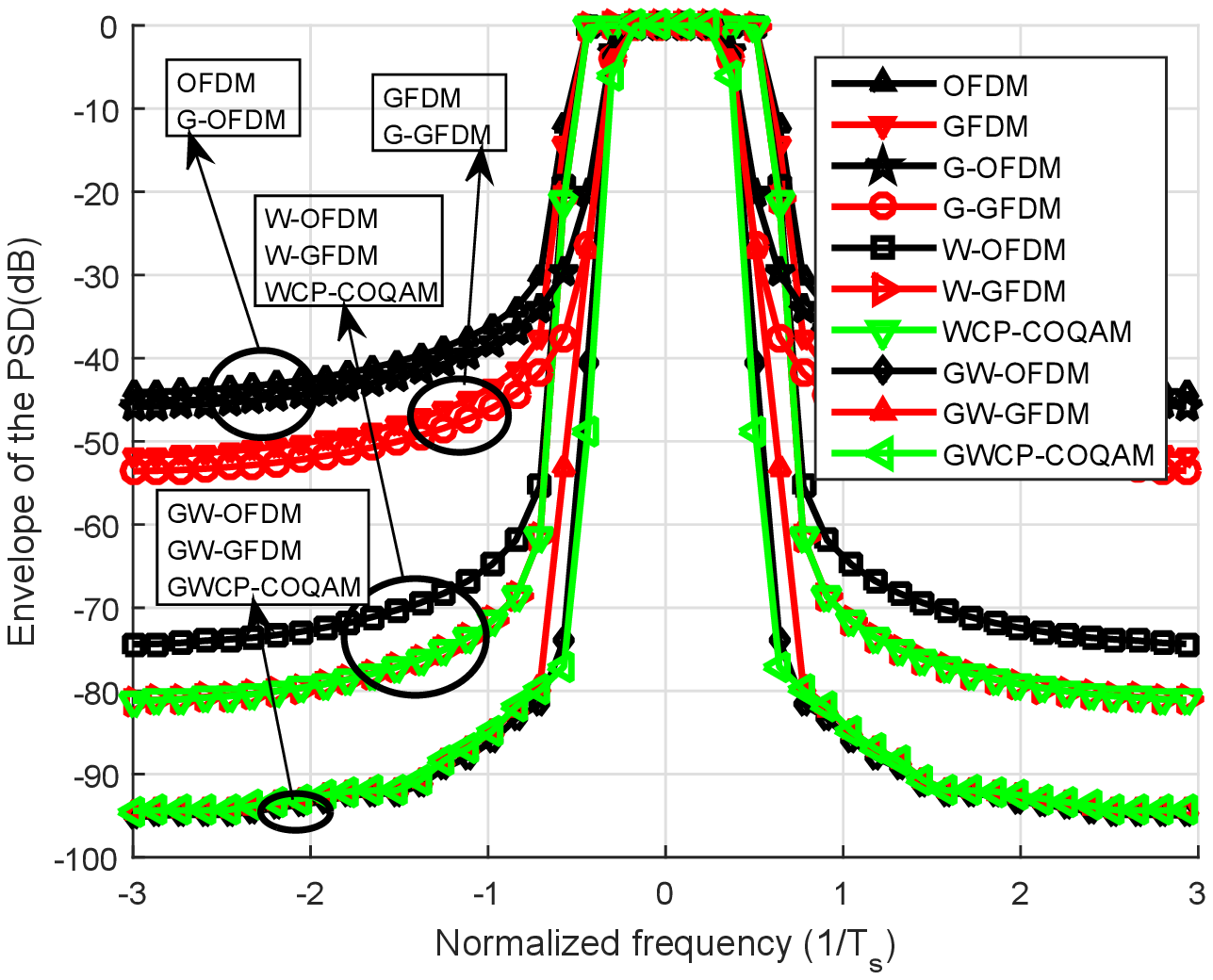}
\caption{\footnotesize PSDs for unequal spectral efficiency when RRC pulse is used with GFDM, WCP-COQAM.}
\label{fig:OOB_1024OFDM}
\end{minipage}
\begin{minipage}[b]{0.42\linewidth}
\graphicspath{ {C:/Bulut/Yuksek_Lisans/Yayinlar/OOB_GFDM/OOB_plots/OOB_plots_v2/} }
\includegraphics[width=\columnwidth]{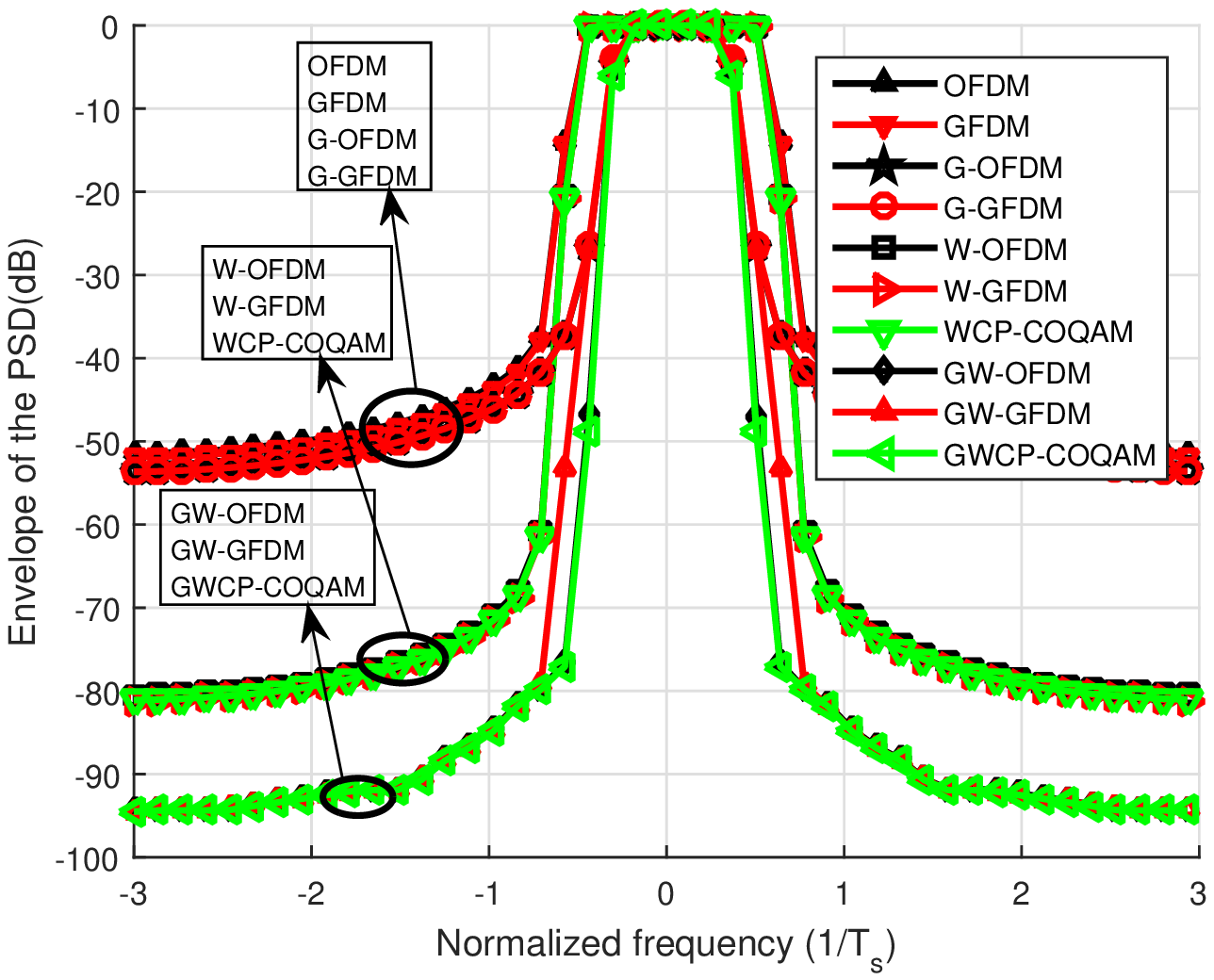}
\caption{\footnotesize PSDs for equal spectral efficiency when RRC pulse is used with GFDM, WCP-COQAM.}
\label{fig:OOB_1024OFDM}
\end{minipage}
\end{figure}
\underline{Error Rate Performance under CFO:}

\begin{figure}[H]
\centering
\graphicspath{ {C:/Bulut/Yuksek_Lisans/Yayinlar/OOB_GFDM/CFO_plots_comm_lett/CFO_plots_v2/} }
\includegraphics[width=0.52\columnwidth]{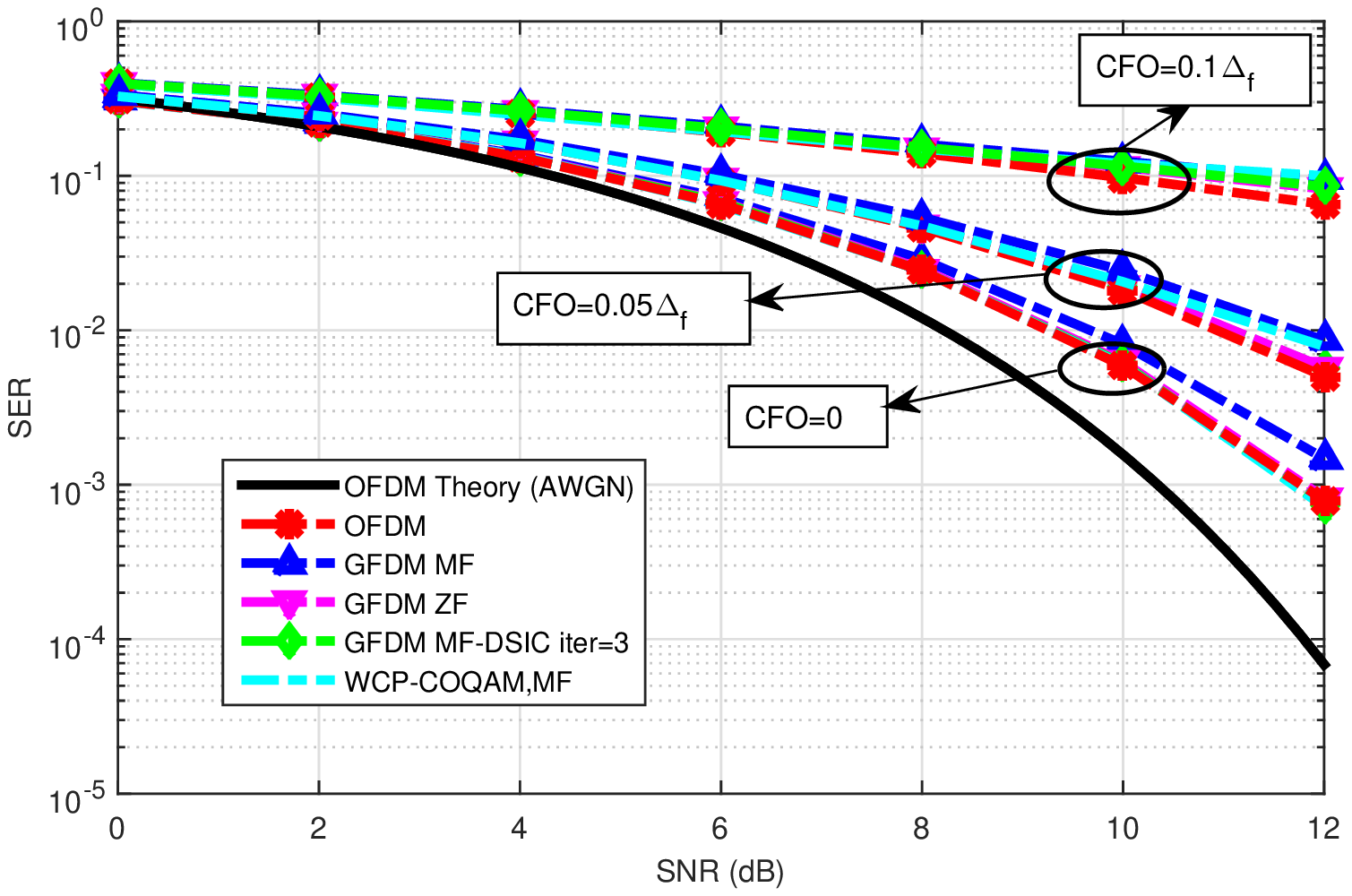}
\caption{Symbol error rate (SER) vs. SNR under CFO for OFDM, GFDM and WCP-COQAM when RRC pulse is used for GFDM and WCP-COQAM.}
\label{fig:OOB_1024OFDM}
\end{figure}
\pagebreak
\subsubsection{\textbf{PHYDYAS pulse}}~\\ \hspace{30pt}\underline{OOB Emissions:}
\begin{figure}[H]
\centering
\begin{minipage}[b]{0.42\linewidth}
\graphicspath{ {C:/Bulut/Yuksek_Lisans/Yayinlar/OOB_GFDM/OOB_plots/OOB_plots_v2/} }
\includegraphics[width=\columnwidth]{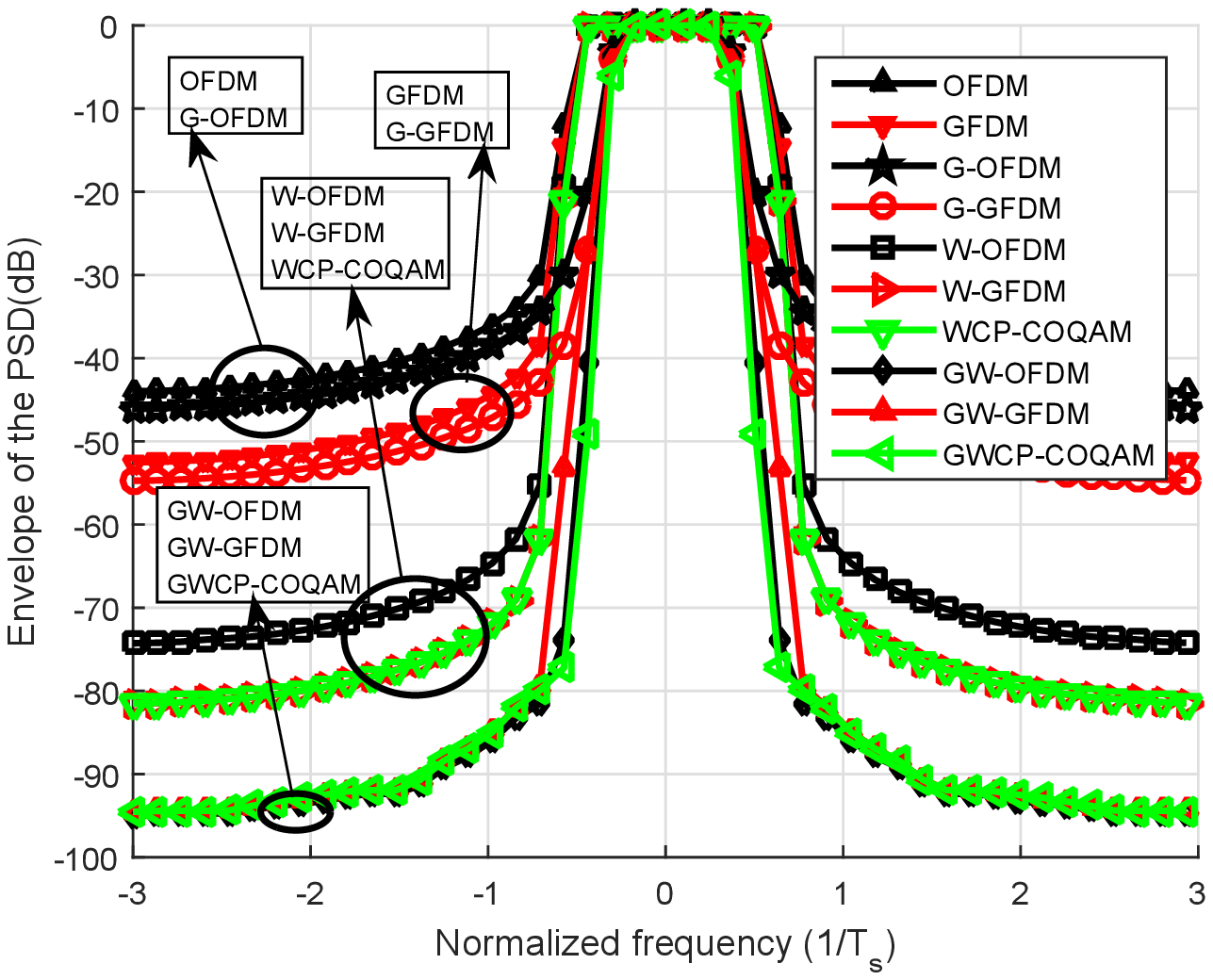}
\caption{\footnotesize PSDs for unequal spectral efficiency when PHYDYAS pulse is used with GFDM, WCP-COQAM.}
\label{fig:OOB_1024OFDM}
\end{minipage}
\begin{minipage}[b]{0.42\linewidth}
\graphicspath{ {C:/Bulut/Yuksek_Lisans/Yayinlar/OOB_GFDM/OOB_plots/OOB_plots_v2/} }
\includegraphics[width=\columnwidth]{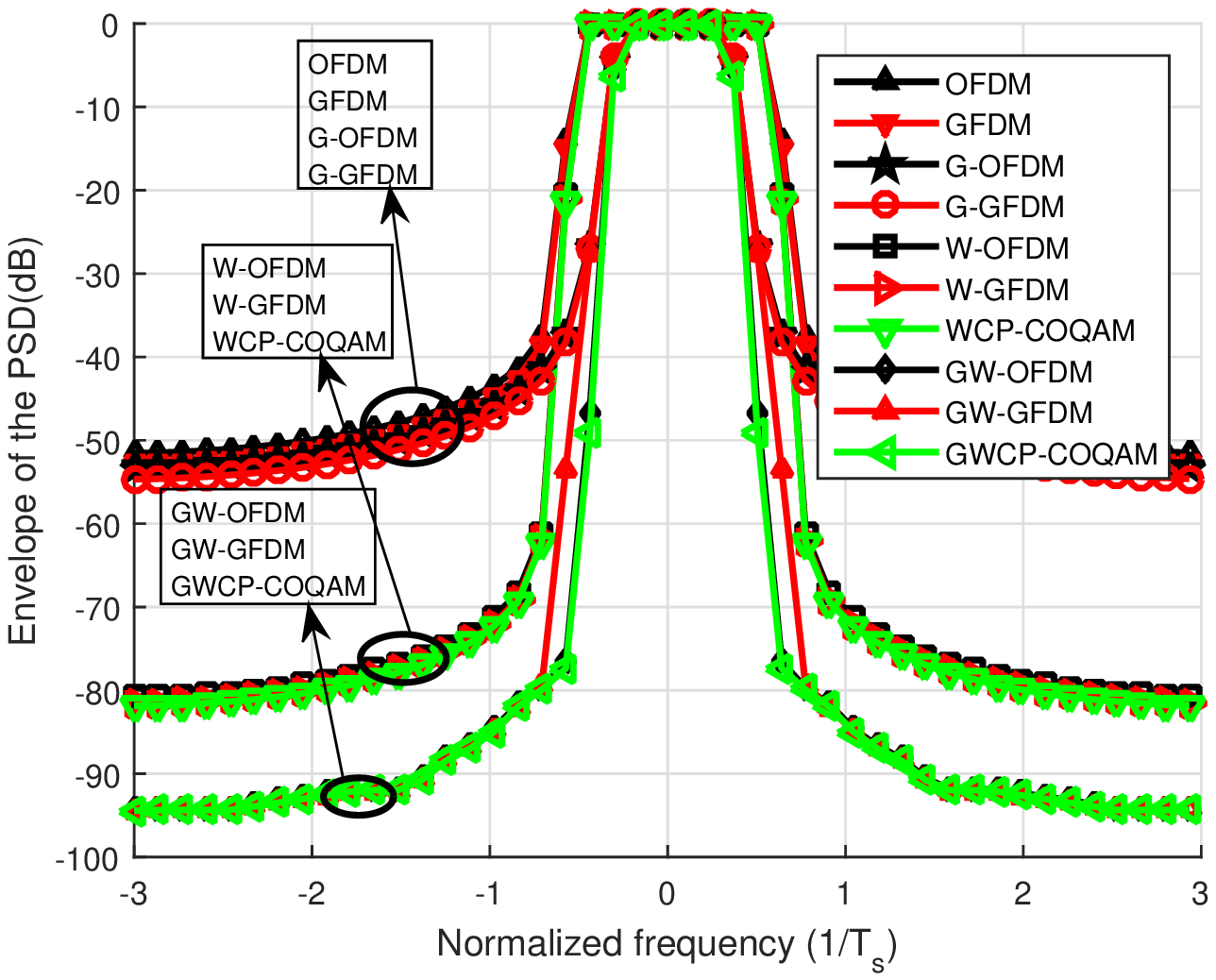}
\caption{\footnotesize PSDs for equal spectral efficiency when PHYDYAS pulse is used with GFDM, WCP-COQAM.}
\label{fig:OOB_1024OFDM}
\end{minipage}
\end{figure}
\underline{Error Rate Performance under CFO:}
\begin{figure}[H]
\centering
\graphicspath{ {C:/Bulut/Yuksek_Lisans/Yayinlar/OOB_GFDM/CFO_plots_comm_lett/CFO_plots_v2/} }
\includegraphics[width=0.55\columnwidth]{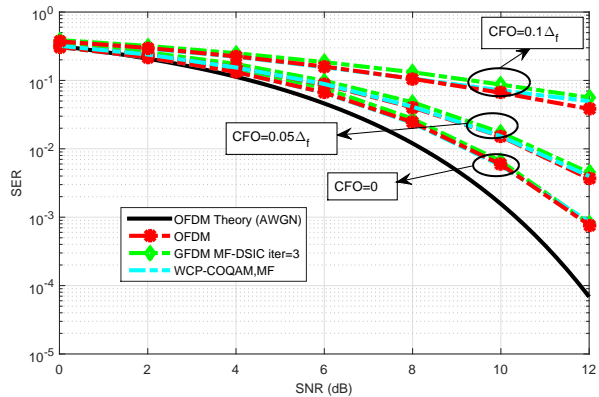}
\caption{Symbol error rate (SER) vs. SNR under CFO for OFDM, GFDM and WCP-COQAM when PHYDYAS pulse is used for GFDM and WCP-COQAM.}
\label{fig:OOB_1024OFDM}
\end{figure}
Only for this part, the number of time slots  ($M$) is selected to be 8 when PHYDYAS pulse is used instead of $M=9$ base scenario value. This value is selected as the closest $M$ value to the $M=9$ base scenario value amongst the possible choices for $M$ specified in \cite{PHYDYAS} for the PHYDYAS pulse.
\pagebreak

\subsubsection{\textbf{IOTA pulse}}~\\
IOTA pulse is generated based on the approximations described in \cite{IOTA}.\\
\underline{OOB Emissions:}
\begin{figure}[H]
\centering
\graphicspath{ {C:/Bulut/Yuksek_Lisans/Yayinlar/OOB_GFDM/OOB_plots/OOB_plots_v2/} }
\includegraphics[width=0.6\columnwidth]{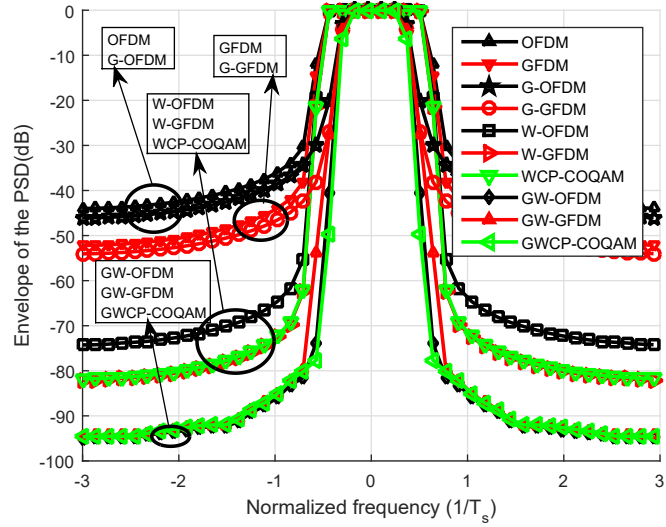}
\caption{PSDs for unequal spectral efficiency when IOTA pulse is used with GFDM, WCP-COQAM.}
\label{fig:OOB_1024OFDM}
\end{figure}
\begin{figure}[H]
\centering
\graphicspath{ {C:/Bulut/Yuksek_Lisans/Yayinlar/OOB_GFDM/OOB_plots/OOB_plots_v2/} }
\includegraphics[width=0.6\columnwidth]{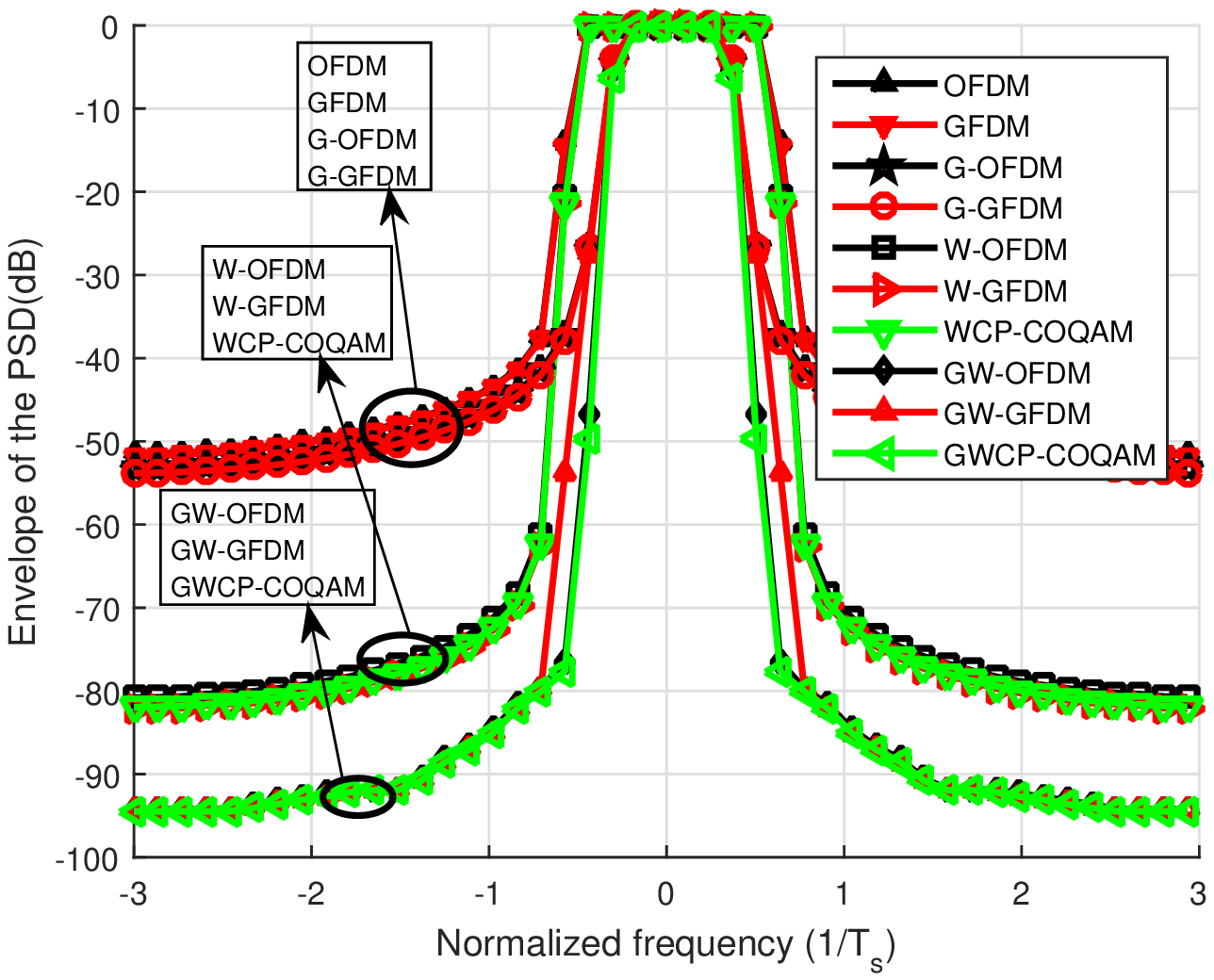}
\caption{PSDs for equal spectral efficiency when IOTA pulse is used with GFDM, WCP-COQAM.}
\label{fig:OOB_1024OFDM}
\end{figure}
\pagebreak

\subsubsection{\textbf{Dirichlet pulse}}
~\\ \hspace{30pt}\underline{OOB Emissions:}
\begin{figure}[H]
\centering
\begin{minipage}[b]{0.49\linewidth}
\graphicspath{ {C:/Bulut/Yuksek_Lisans/Yayinlar/OOB_GFDM/OOB_plots/OOB_plots_v2/} }
\includegraphics[width=\columnwidth]{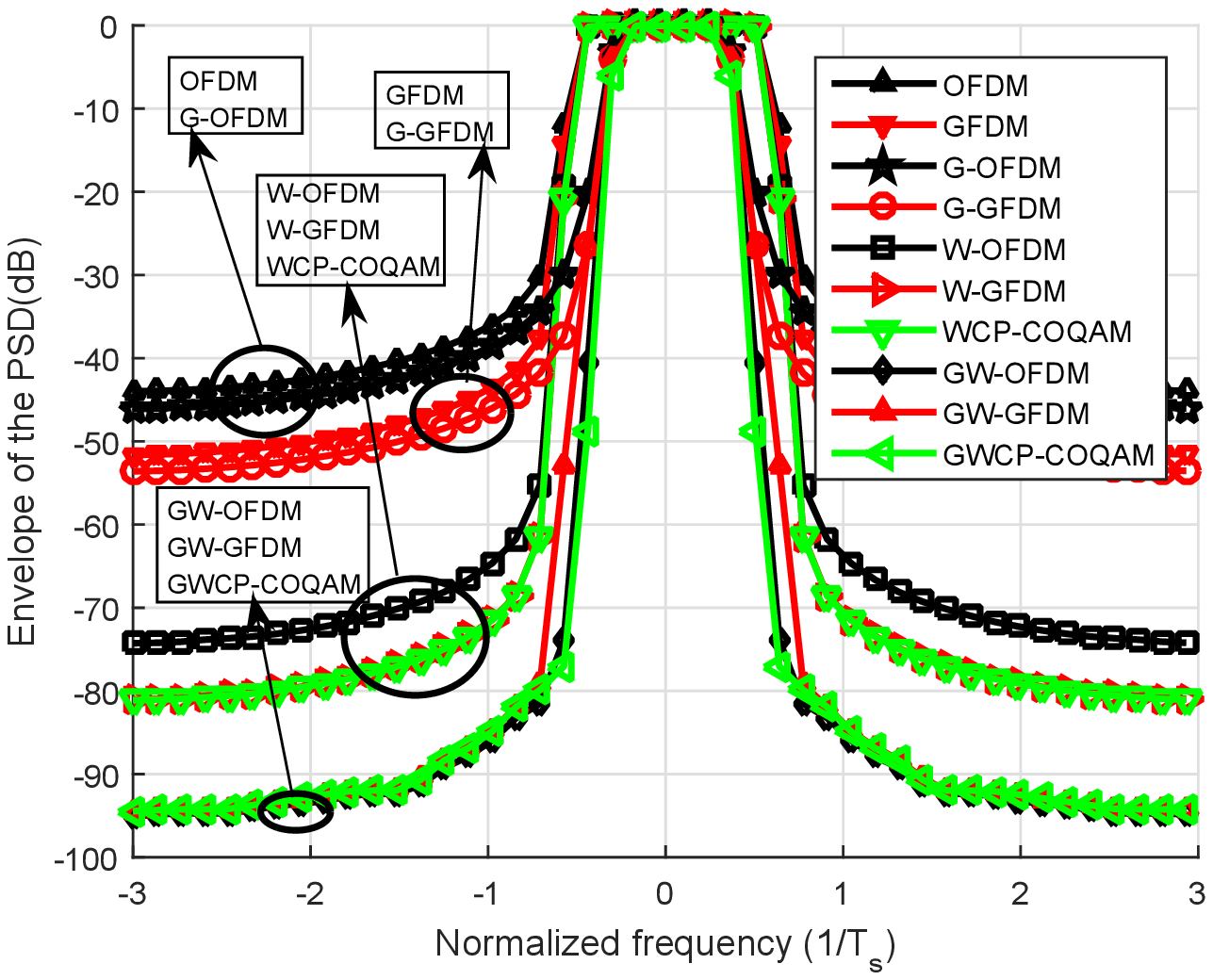}
\caption{\footnotesize PSDs for unequal spectral efficiency when Dirichlet pulse is used with GFDM, WCP-COQAM.}
\label{fig:OOB_1024OFDM}
\end{minipage}
\begin{minipage}[b]{0.49\linewidth}
\graphicspath{ {C:/Bulut/Yuksek_Lisans/Yayinlar/OOB_GFDM/OOB_plots/OOB_plots_v2/} }
\includegraphics[width=\columnwidth]{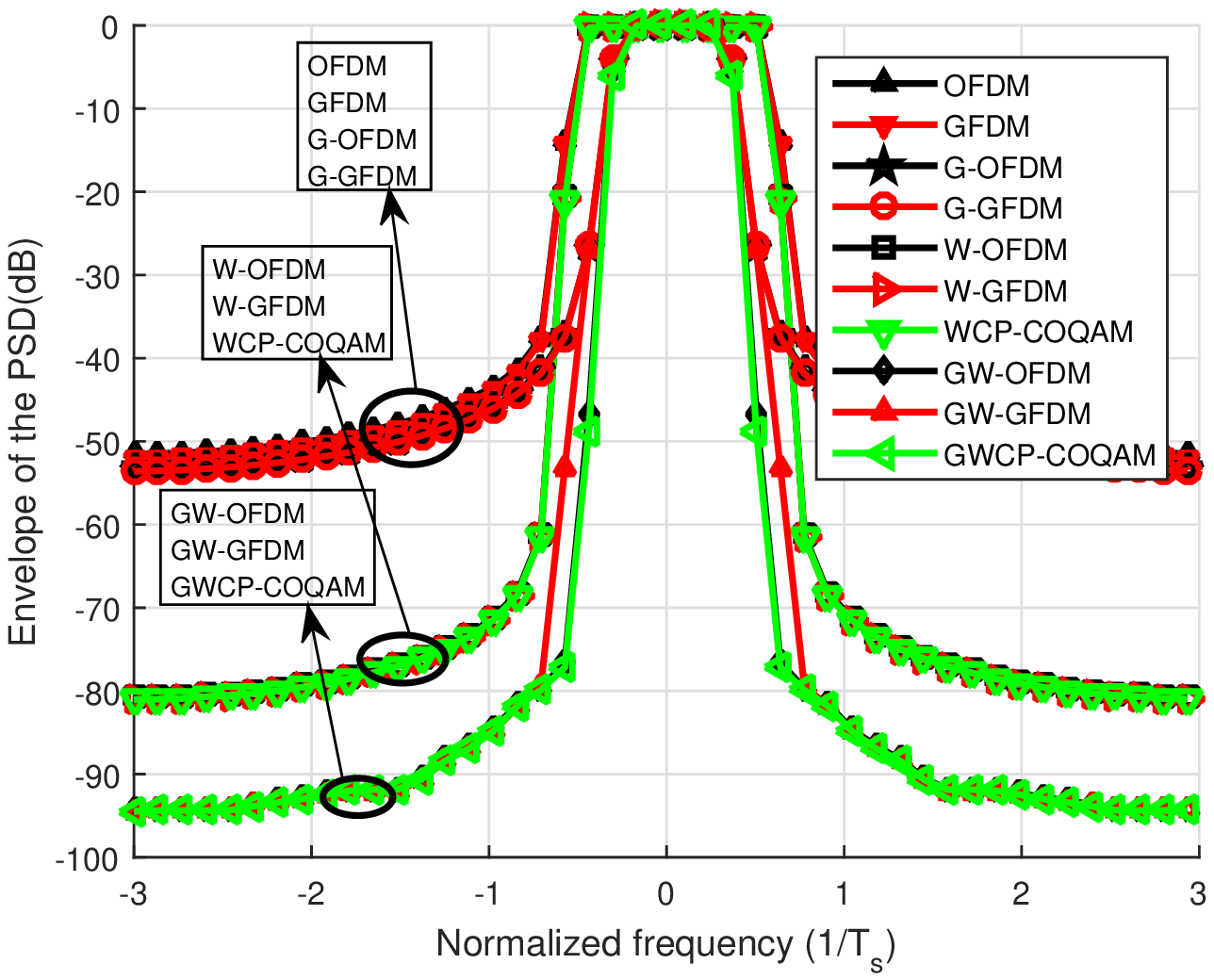}
\caption{\footnotesize PSDs for equal spectral efficiency when Dirichlet pulse is used with GFDM, WCP-COQAM.}
\label{fig:OOB_1024OFDM}
\end{minipage}
\end{figure}
\underline{Error Rate Performance under CFO:}

\begin{figure}[H]
\centering
\graphicspath{ {C:/Bulut/Yuksek_Lisans/Yayinlar/OOB_GFDM/CFO_plots_comm_lett/CFO_plots_v2/} }
\includegraphics[width=0.6\columnwidth]{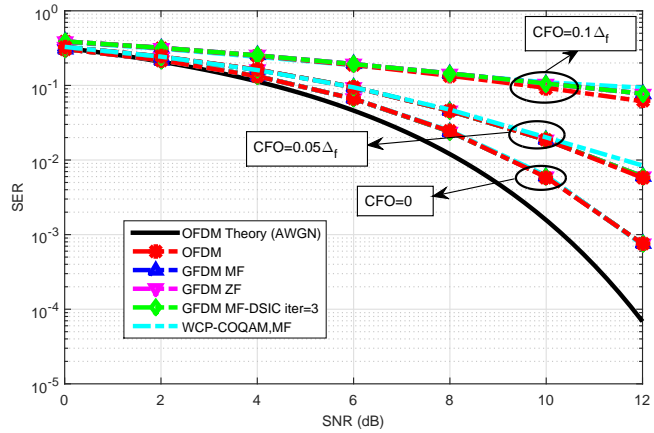}
\caption{Symbol error rate (SER) vs. SNR under CFO for OFDM, GFDM and WCP-COQAM when Dirichlet pulse is used for GFDM and WCP-COQAM.}
\label{fig:OOB_1024OFDM}
\end{figure}
\pagebreak
\subsection{\textbf{Variations in the Number of Subcarriers (K) and Time Slots (M)}}
In this section, M and K values in the base scenario (M=9, K=128) will be changed to obtain OOB emisssions for equal and unequal spectral efficiency cases. The error rate performances under CFO will also be presented.

\subsubsection{\textbf{M=5, K=128}}
~\\ \hspace{30pt}\underline{OOB Emissions:}
\begin{figure}[H]
\centering
\begin{minipage}[b]{0.42\linewidth}
\graphicspath{ {C:/Bulut/Yuksek_Lisans/Yayinlar/OOB_GFDM/OOB_plots/OOB_plots_v2/} }
\includegraphics[width=\columnwidth]{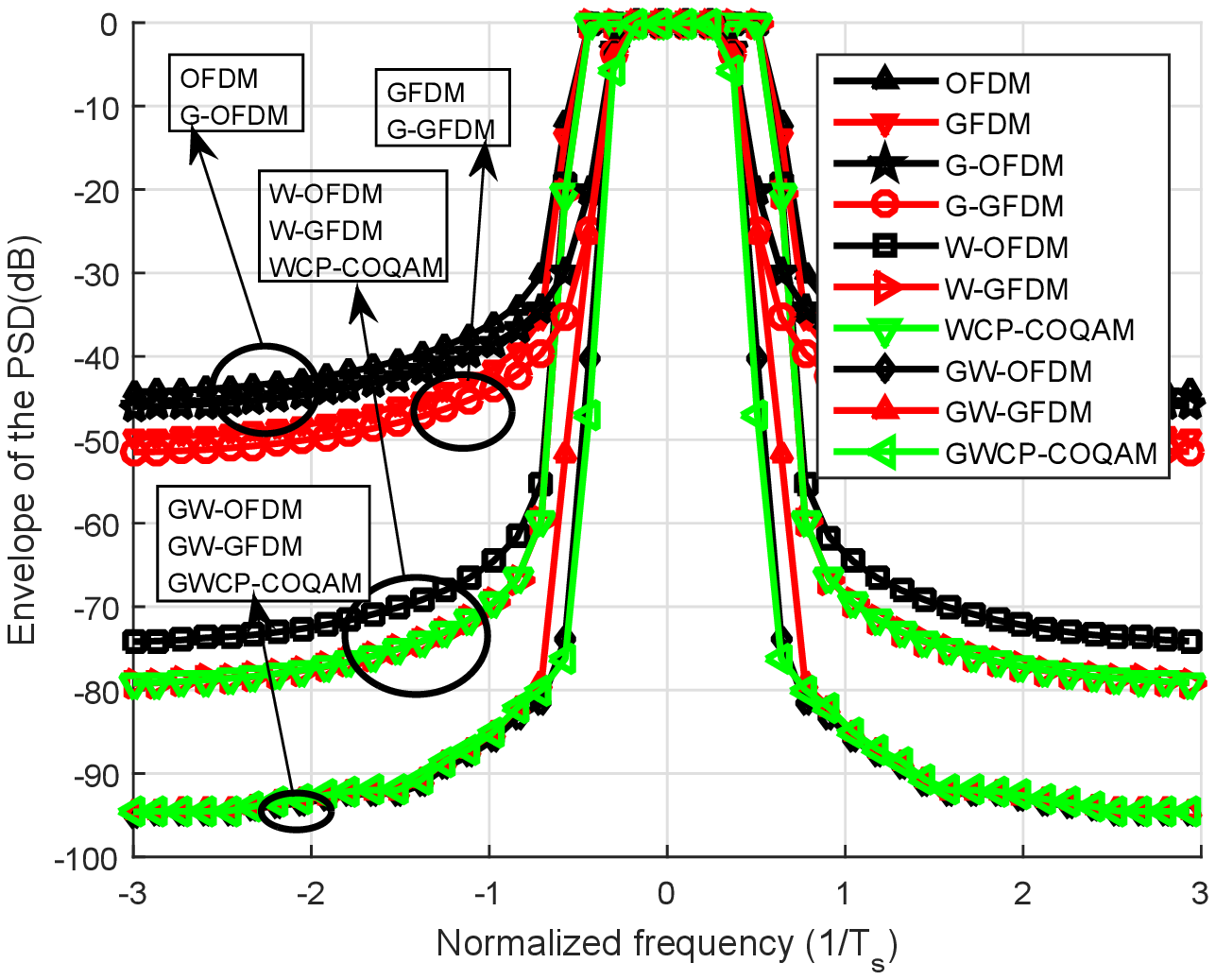}
\caption{PSDs for unequal spectral efficiency when K=128 and M=5.}
\label{fig:OOB_1024OFDM}
\end{minipage}
\begin{minipage}[b]{0.42\linewidth}
\graphicspath{ {C:/Bulut/Yuksek_Lisans/Yayinlar/OOB_GFDM/OOB_plots/OOB_plots_v2/} }
\includegraphics[width=\columnwidth]{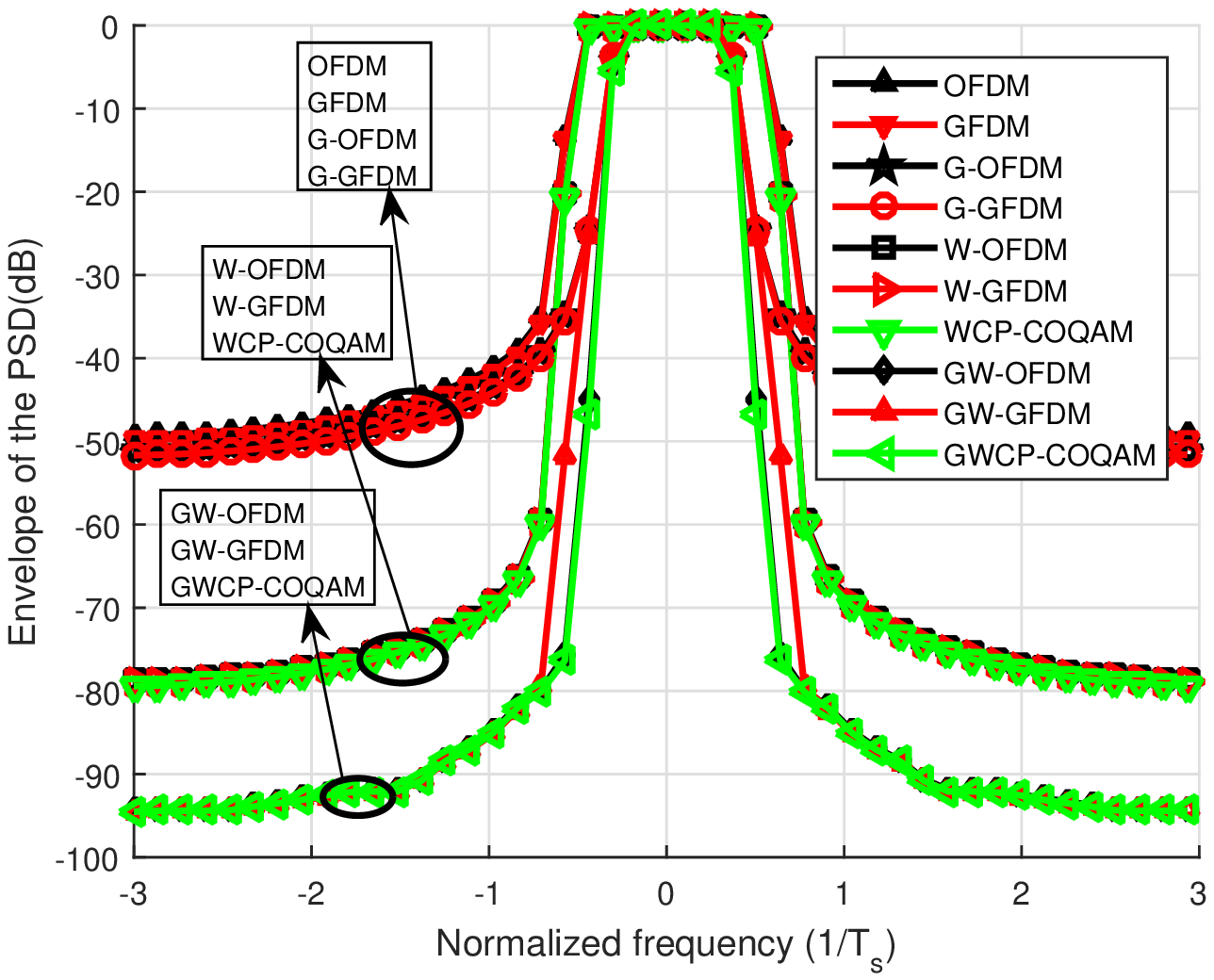}
\caption{PSDs for equal spectral efficiency when K=128 and M=5.}
\label{fig:OOB_1024OFDM}
\end{minipage}
\end{figure}
\underline{Error Rate Performance under CFO:}

\begin{figure}[H]
\centering
\graphicspath{ {C:/Bulut/Yuksek_Lisans/Yayinlar/OOB_GFDM/CFO_plots_comm_lett/CFO_plots_v2/} }
\includegraphics[width=0.56\columnwidth]{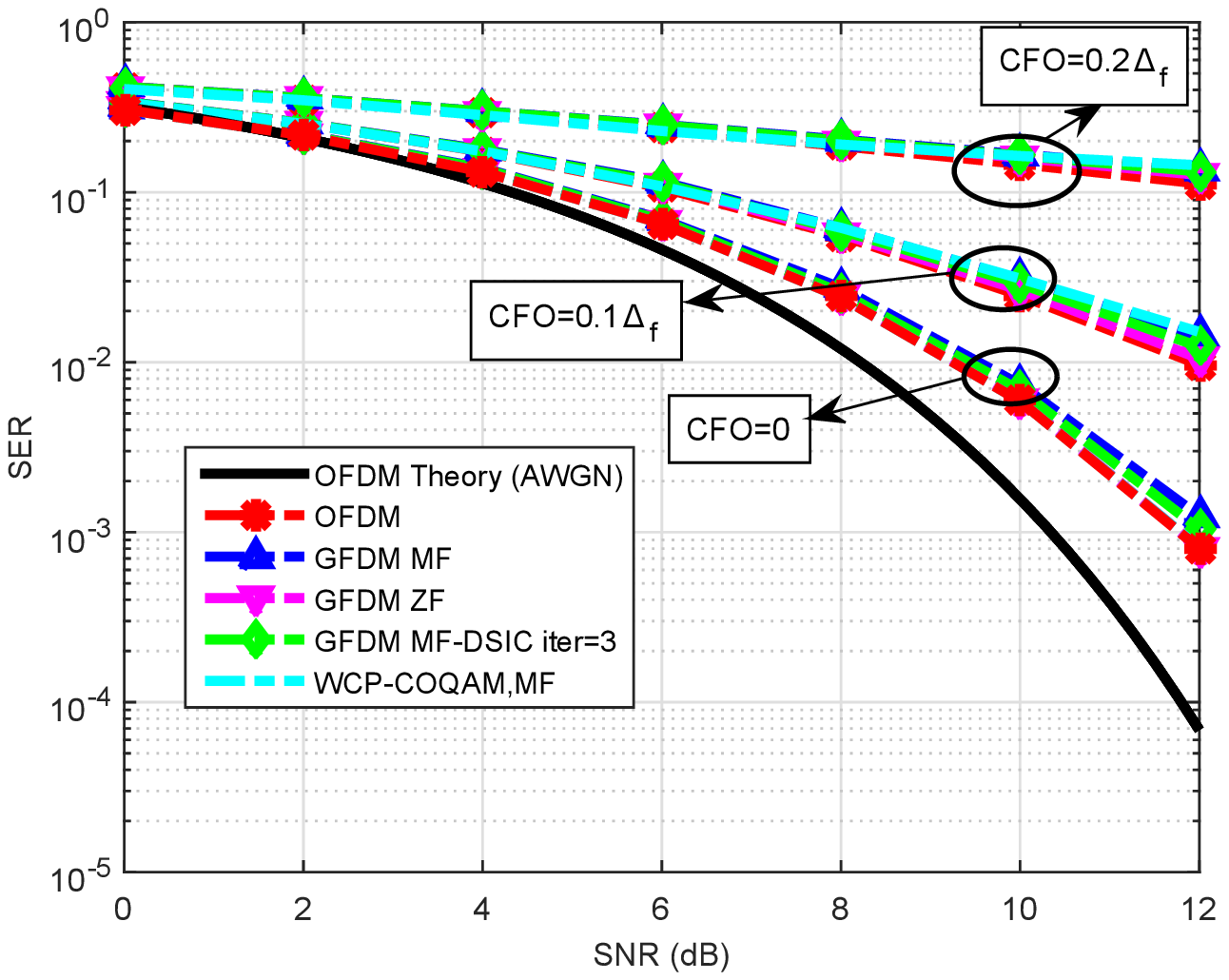}
\caption{Symbol error rate (SER) vs. SNR under CFO for OFDM, GFDM and WCP-COQAM for K=128 and M=5.}
\label{fig:OOB_1024OFDM}
\end{figure}

\pagebreak

\subsubsection{\textbf{M=5, K=64}}
~\\ \hspace{30pt}\underline{OOB Emissions:}
\begin{figure}[H]
\centering
\begin{minipage}[b]{0.49\linewidth}
\graphicspath{ {C:/Bulut/Yuksek_Lisans/Yayinlar/OOB_GFDM/OOB_plots/OOB_plots_v2/} }
\includegraphics[width=\columnwidth]{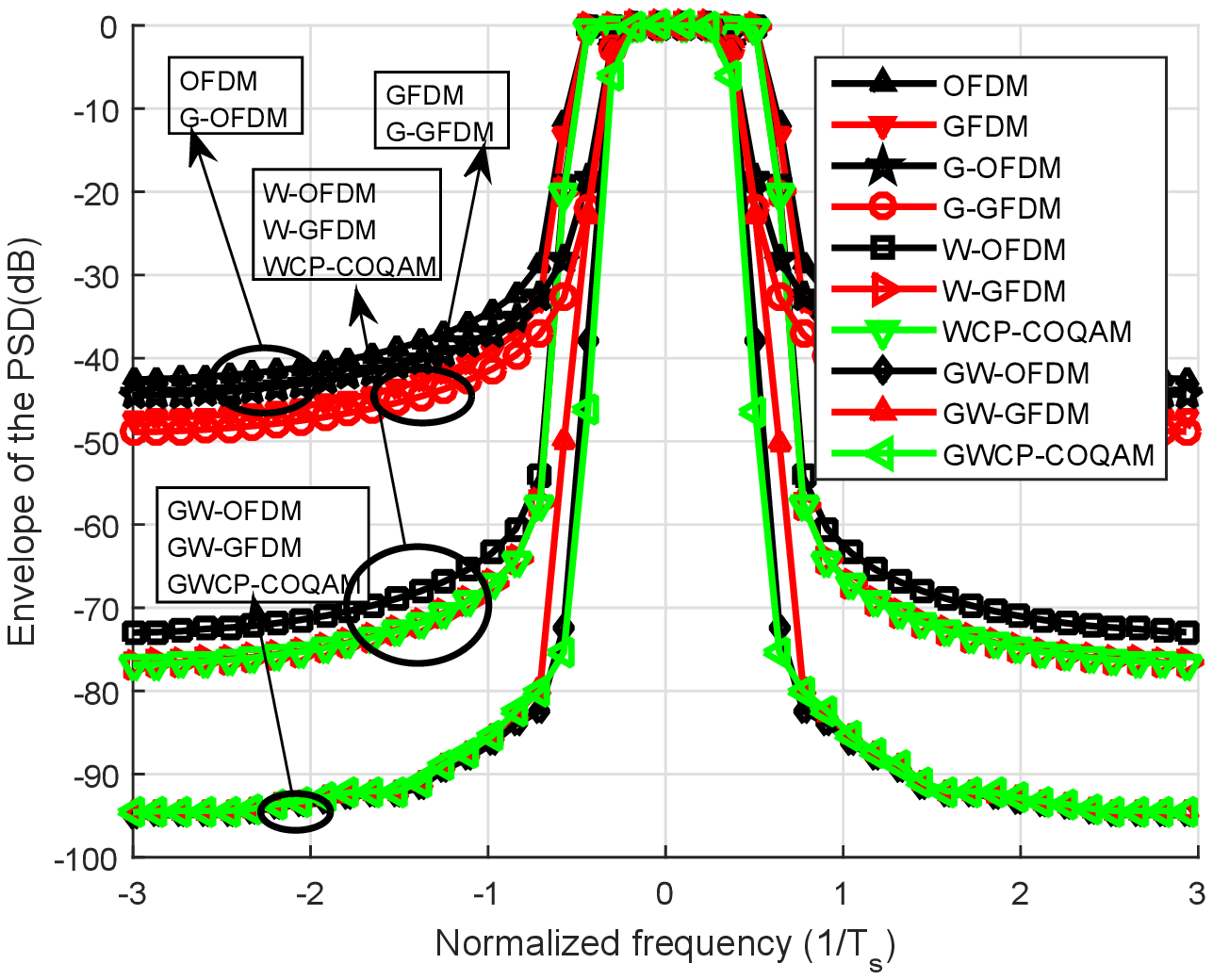}
\caption{\footnotesize PSDs for unequal spectral efficiency when K=64 and M=5.}
\label{fig:OOB_1024OFDM}
\end{minipage}
\begin{minipage}[b]{0.49\linewidth}
\graphicspath{ {C:/Bulut/Yuksek_Lisans/Yayinlar/OOB_GFDM/OOB_plots/OOB_plots_v2/} }
\includegraphics[width=\columnwidth]{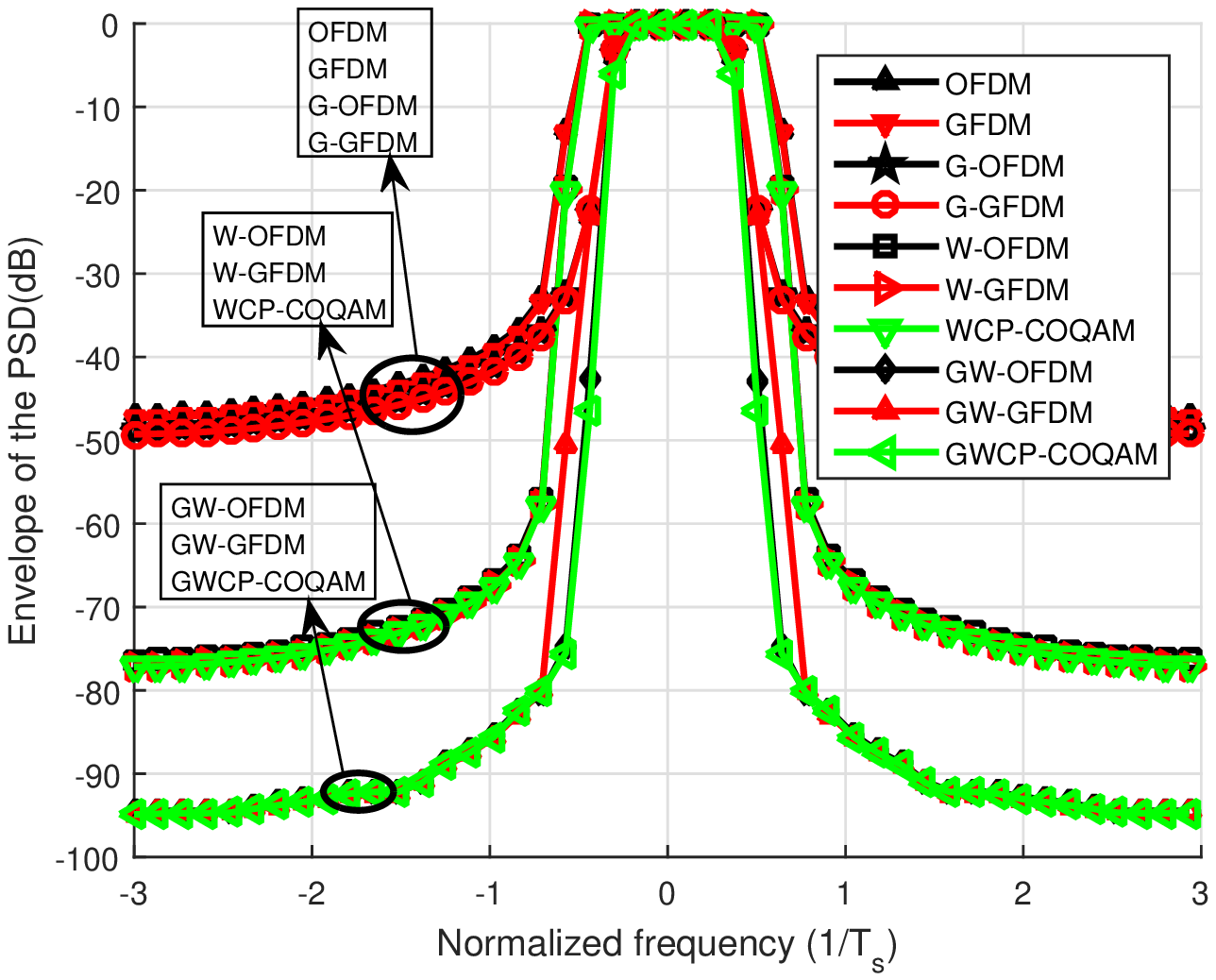}
\caption{\footnotesize PSDs for equal spectral efficiency when K=64 and M=5.}
\label{fig:OOB_1024OFDM}
\end{minipage}
\end{figure}
\underline{Error Rate Performance under CFO:}

\begin{figure}[H]
\centering
\graphicspath{ {C:/Bulut/Yuksek_Lisans/Yayinlar/OOB_GFDM/CFO_plots_comm_lett/CFO_plots_v2/} }
\includegraphics[width=0.6\columnwidth]{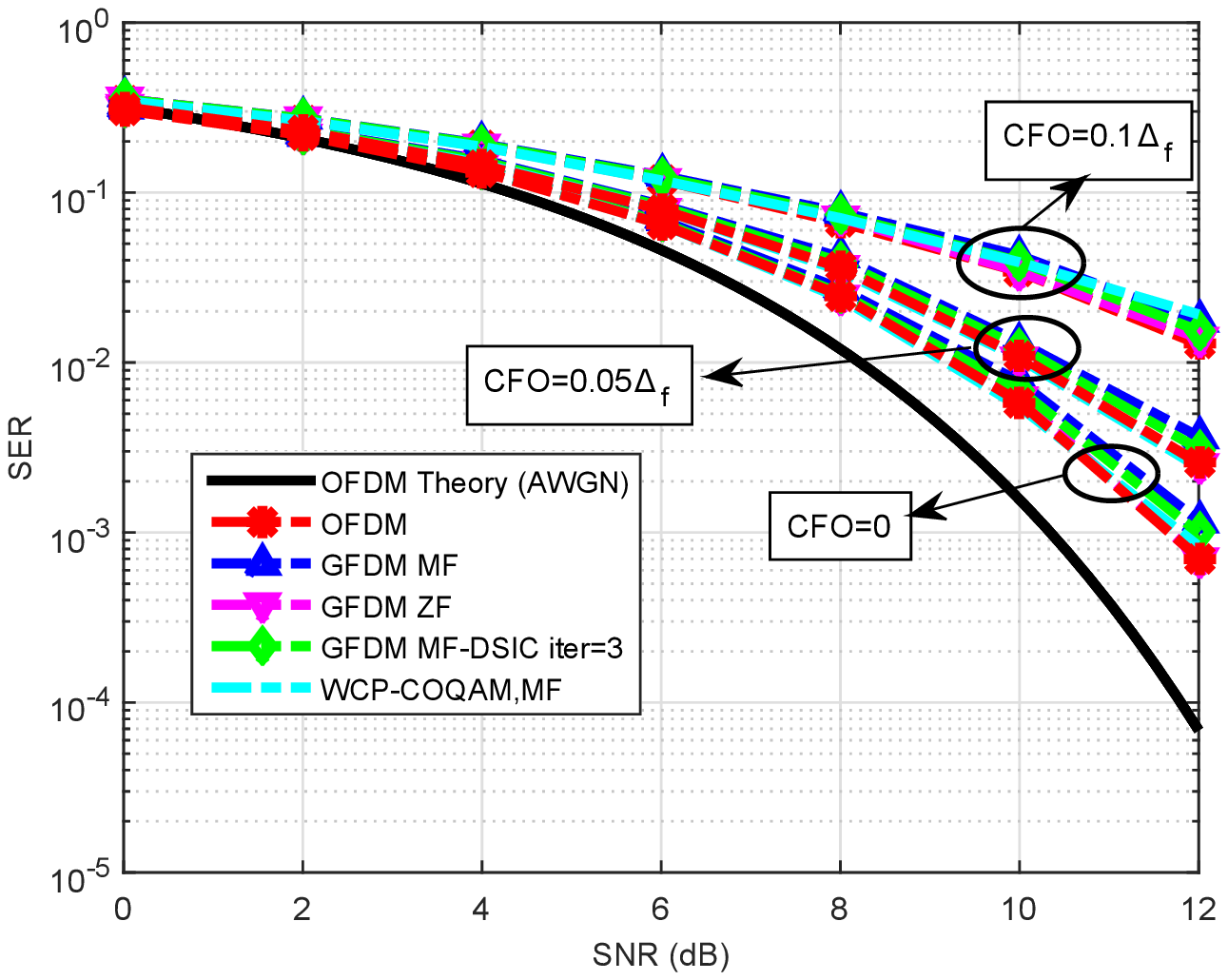}
\caption{Symbol error rate (SER) vs. SNR under CFO for OFDM, GFDM and WCP-COQAM for K=64 and M=5.}
\label{fig:OOB_1024OFDM}
\end{figure}

\pagebreak

\subsubsection{\textbf{M=9, K=64}}
~\\ \hspace{30pt}\underline{OOB Emissions:}
\begin{figure}[H]
\centering
\begin{minipage}[b]{0.49\linewidth}
\graphicspath{ {C:/Bulut/Yuksek_Lisans/Yayinlar/OOB_GFDM/OOB_plots/OOB_plots_v2/} }
\includegraphics[width=\columnwidth]{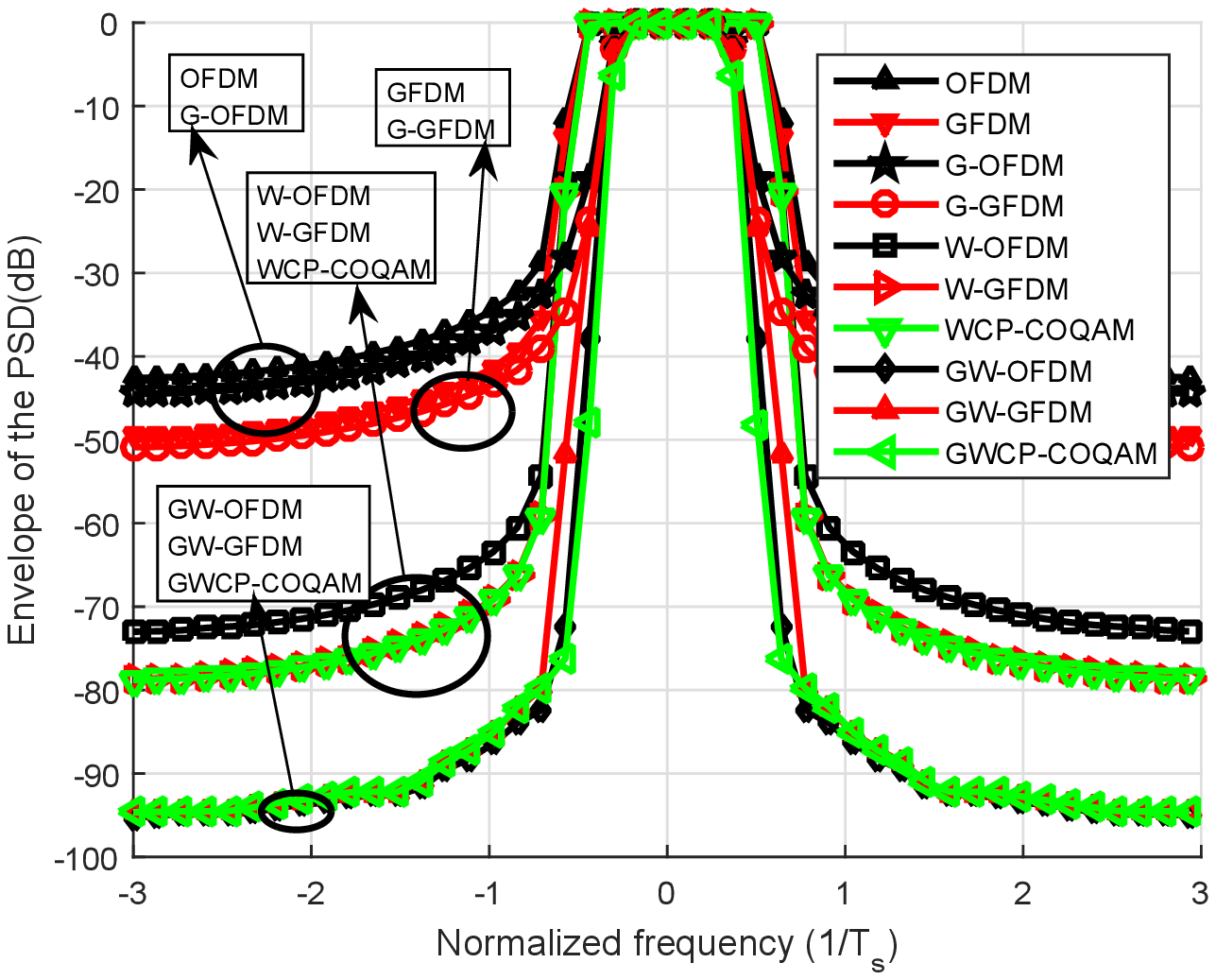}
\caption{\footnotesize PSDs for unequal spectral efficiency when K=64 and M=9.}
\label{fig:OOB_1024OFDM}
\end{minipage}
\begin{minipage}[b]{0.49\linewidth}
\graphicspath{ {C:/Bulut/Yuksek_Lisans/Yayinlar/OOB_GFDM/OOB_plots/OOB_plots_v2/} }
\includegraphics[width=\columnwidth]{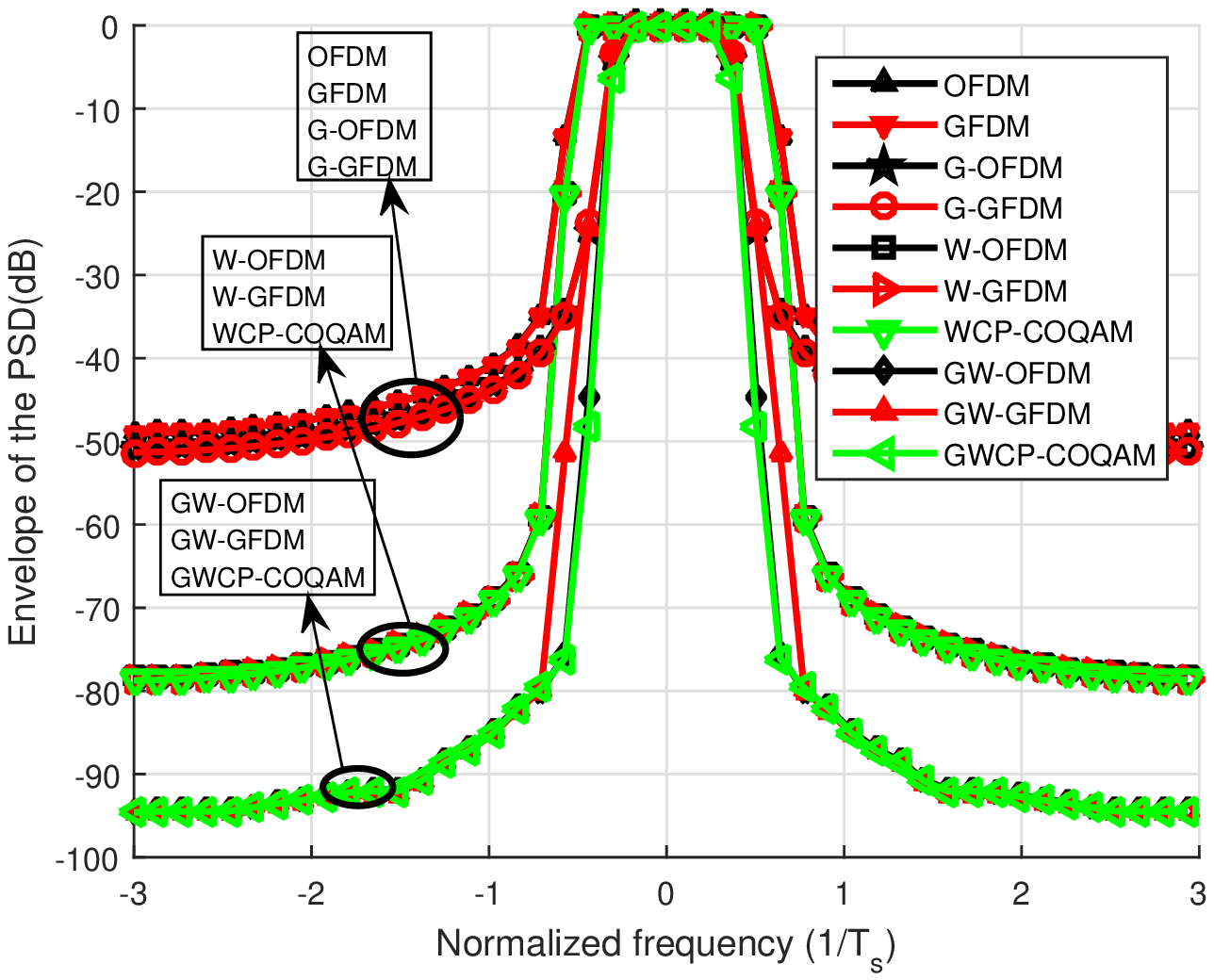}
\caption{\footnotesize PSDs for equal spectral efficiency when K=64 and M=9.}
\label{fig:OOB_1024OFDM}
\end{minipage}
\end{figure}
\underline{Error Rate Performance under CFO:}

\begin{figure}[H]
\centering
\graphicspath{ {C:/Bulut/Yuksek_Lisans/Yayinlar/OOB_GFDM/CFO_plots_comm_lett/CFO_plots_v2/} }
\includegraphics[width=0.6\columnwidth]{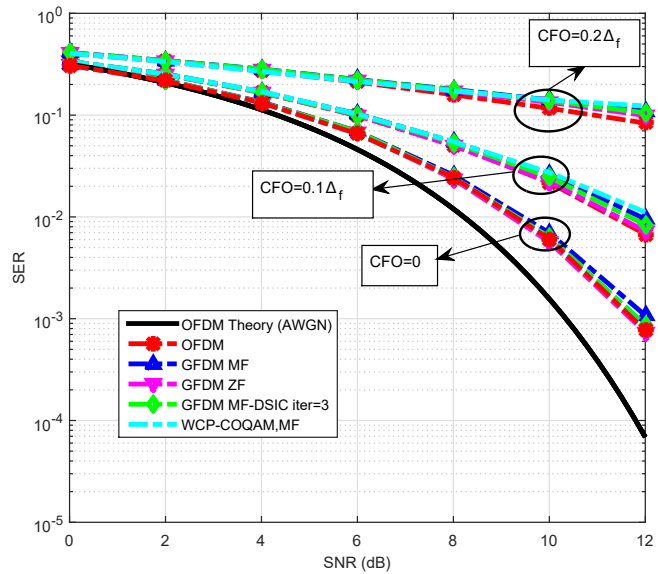}
\caption{Symbol error rate (SER) vs. SNR under CFO for OFDM, GFDM and WCP-COQAM for K=64 and M=9.}
\label{fig:OOB_1024OFDM}
\end{figure}
\pagebreak
\subsection{\textbf{Variations in the Window Length, Oversampling (Interpolation) Filter Length, Cyclic Prefix Length, Oversampling Rate, Roll-off Factor of RC Pulse and Constellation Size}}
In this section, window length, cyclic prefix (CP) length, roll-off factor of the RC pulse used for GFDM or WCP-COQAM and the constellation size in the base scenario (which are 18 samples from both sides of OFDM/GFDM/WCP-COQAM symbols, 32 samples, 0.1 and 4-QAM, respectively) will be changed to obtain OOB emisssions for equal and unequal spectral efficiency cases. The error rate performances under CFO will also be presented. \pagebreak
\subsubsection{\textbf{Window Length=36}}
~\\ \hspace{30pt}\underline{OOB Emissions:}
\begin{figure}[H]
\centering
\begin{minipage}[b]{0.49\linewidth}
\graphicspath{ {C:/Bulut/Yuksek_Lisans/Yayinlar/OOB_GFDM/OOB_plots/OOB_plots_v2/} }
\includegraphics[width=\columnwidth]{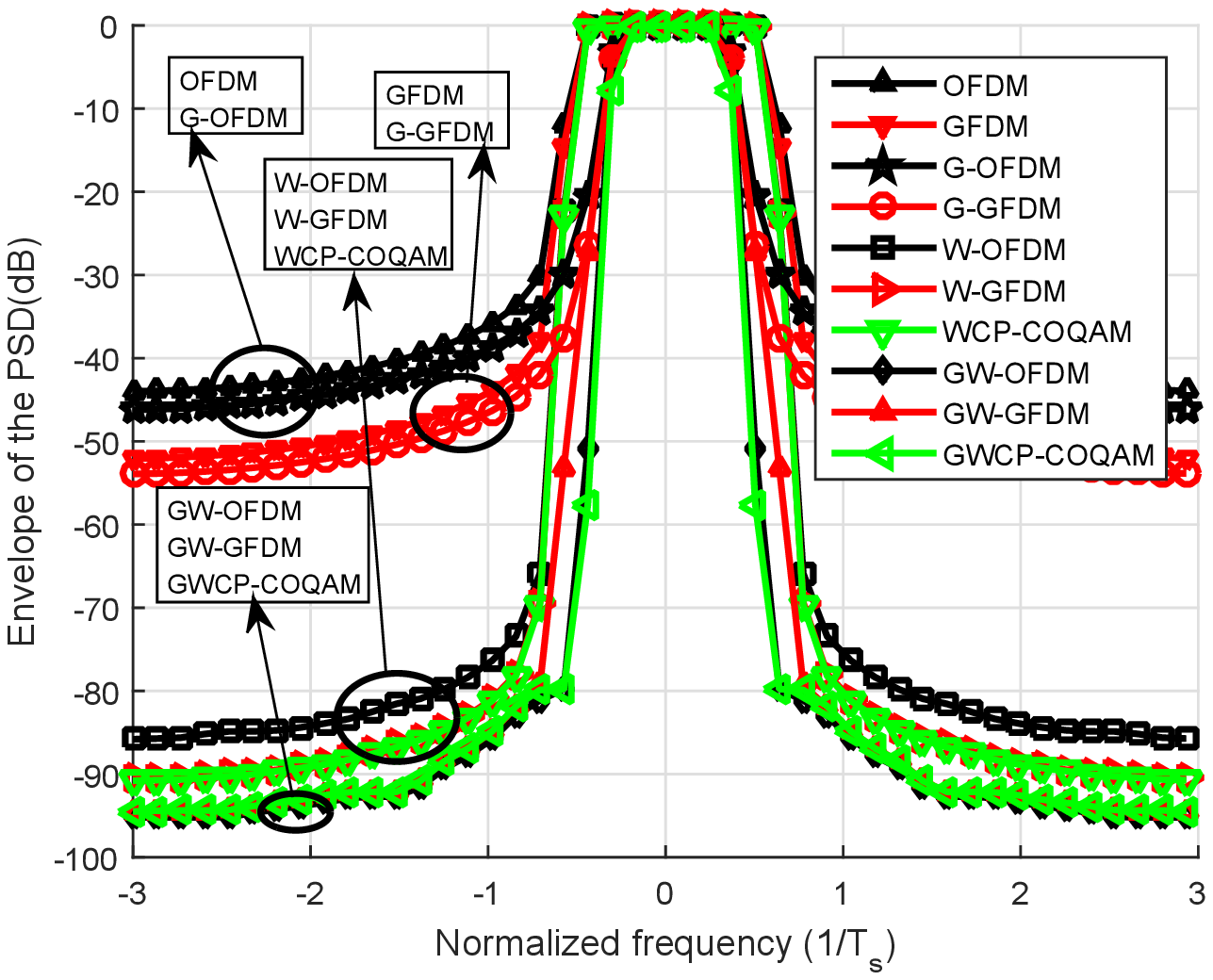}
\caption{\footnotesize PSDs for unequal spectral efficiency when window length=36 samples.}
\label{fig:OOB_1024OFDM}
\end{minipage}
\begin{minipage}[b]{0.49\linewidth}
\graphicspath{ {C:/Bulut/Yuksek_Lisans/Yayinlar/OOB_GFDM/OOB_plots/OOB_plots_v2/} }
\includegraphics[width=\columnwidth]{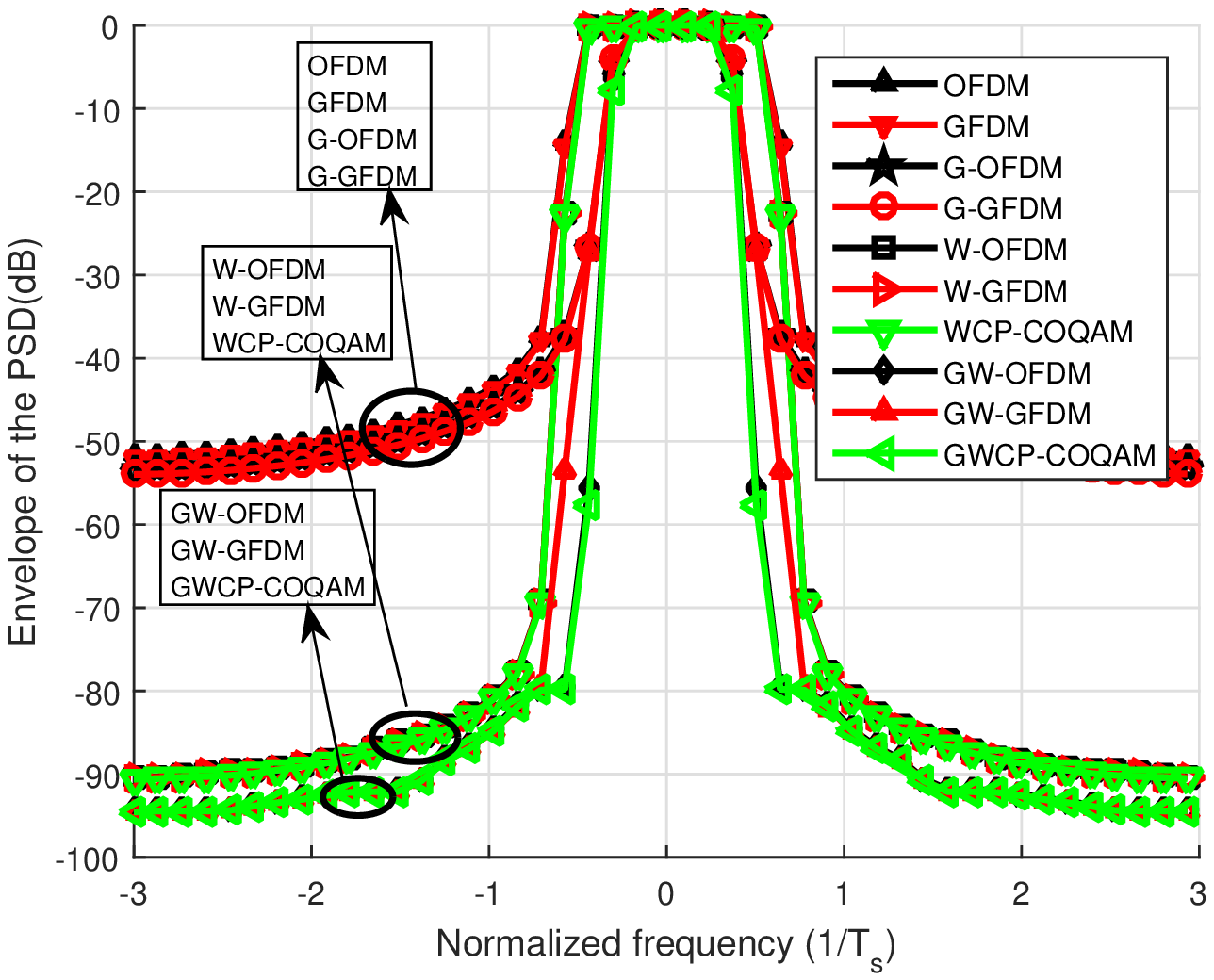}
\caption{\footnotesize PSDs for equal spectral efficiency when window length=36 samples.}
\label{fig:OOB_1024OFDM}
\end{minipage}
\end{figure}
\underline{Error Rate Performance under CFO:}

\begin{figure}[H]
\centering
\graphicspath{ {C:/Bulut/Yuksek_Lisans/Yayinlar/OOB_GFDM/CFO_plots_comm_lett/CFO_plots_v2/} }
\includegraphics[width=0.6\columnwidth]{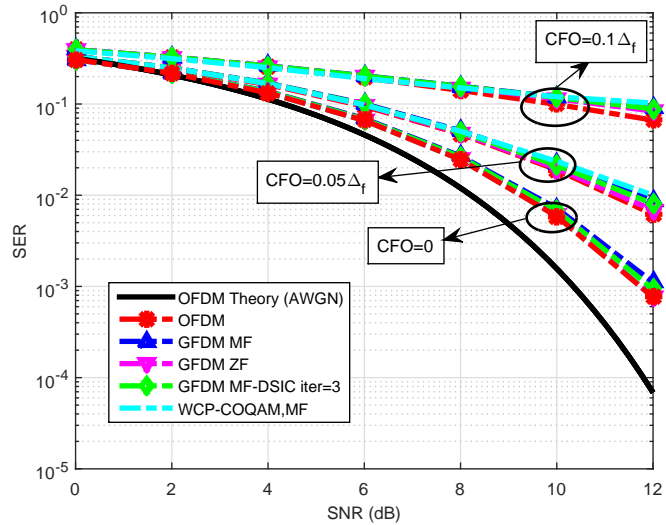}
\caption{Symbol error rate (SER) vs. SNR under CFO for OFDM, GFDM and WCP-COQAM when window length=36 samples.}
\label{fig:OOB_1024OFDM}
\end{figure}

\pagebreak
\subsubsection{\textbf{Oversampling Filter Length=41}}
~\\ \hspace{30pt}\underline{OOB Emissions:}

\begin{figure}[H]
\centering
\graphicspath{ {C:/Bulut/Yuksek_Lisans/Yayinlar/OOB_GFDM/OOB_plots/OOB_plots_v2/} }
\includegraphics[width=0.6\columnwidth]{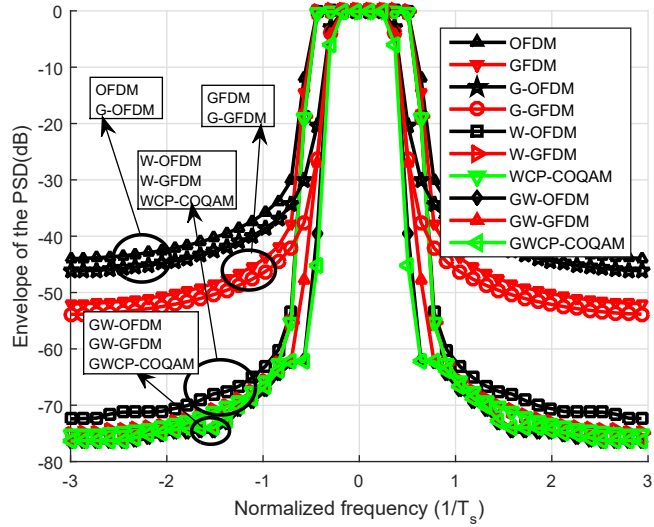}
\caption{PSDs for unequal spectral efficiency, non-contiguous case}
\label{fig:noncontig_uneq}
\end{figure}

\begin{figure}[H]
\centering
\graphicspath{ {C:/Bulut/Yuksek_Lisans/Yayinlar/OOB_GFDM/OOB_plots/OOB_plots_v2/} }
\includegraphics[width=0.6\columnwidth]{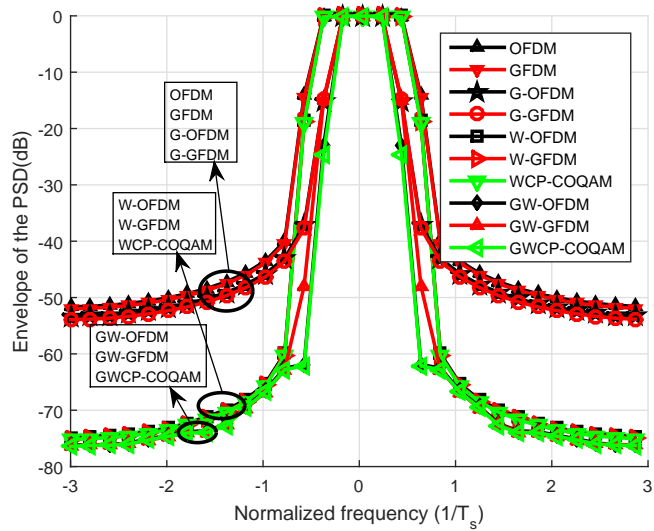}
\caption{PSDs for equal spectral efficiency, non-contiguous case}
\label{fig:contig_uneq}
\end{figure}

Here no CFO error rate curve is included since oversampling filter is only affects OOB spectrum.

\pagebreak
\subsubsection{\textbf{CP Length=64}}
~\\ \hspace{30pt}\underline{OOB Emissions:}
\begin{figure}[H]
\centering
\begin{minipage}[b]{0.49\linewidth}
\graphicspath{ {C:/Bulut/Yuksek_Lisans/Yayinlar/OOB_GFDM/OOB_plots/OOB_plots_v2/} }
\includegraphics[width=\columnwidth]{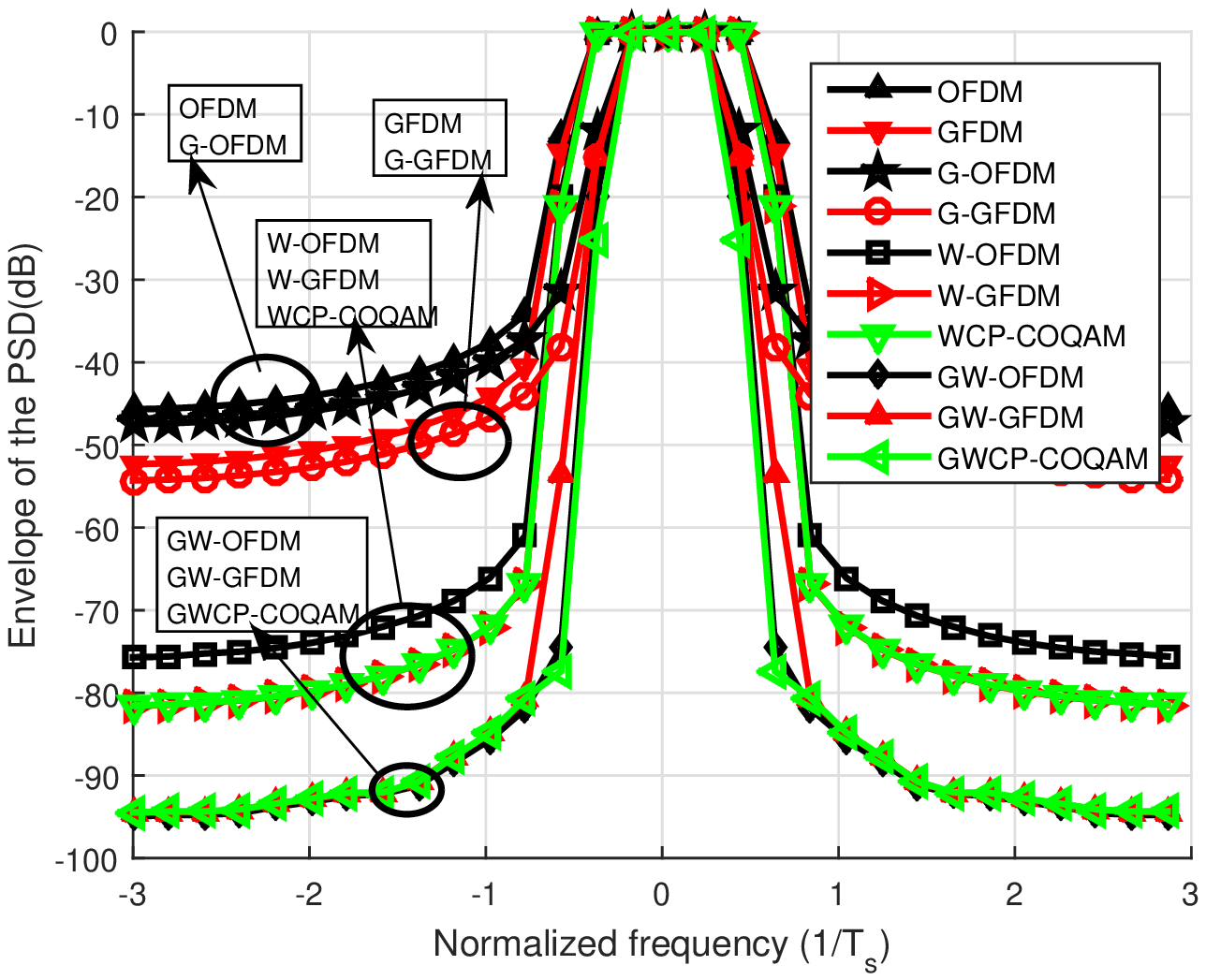}
\caption{\footnotesize PSDs for unequal spectral efficiency when CP length=64 samples.}
\label{fig:OOB_1024OFDM}
\end{minipage}
\begin{minipage}[b]{0.49\linewidth}
\graphicspath{ {C:/Bulut/Yuksek_Lisans/Yayinlar/OOB_GFDM/OOB_plots/OOB_plots_v2/} }
\includegraphics[width=\columnwidth]{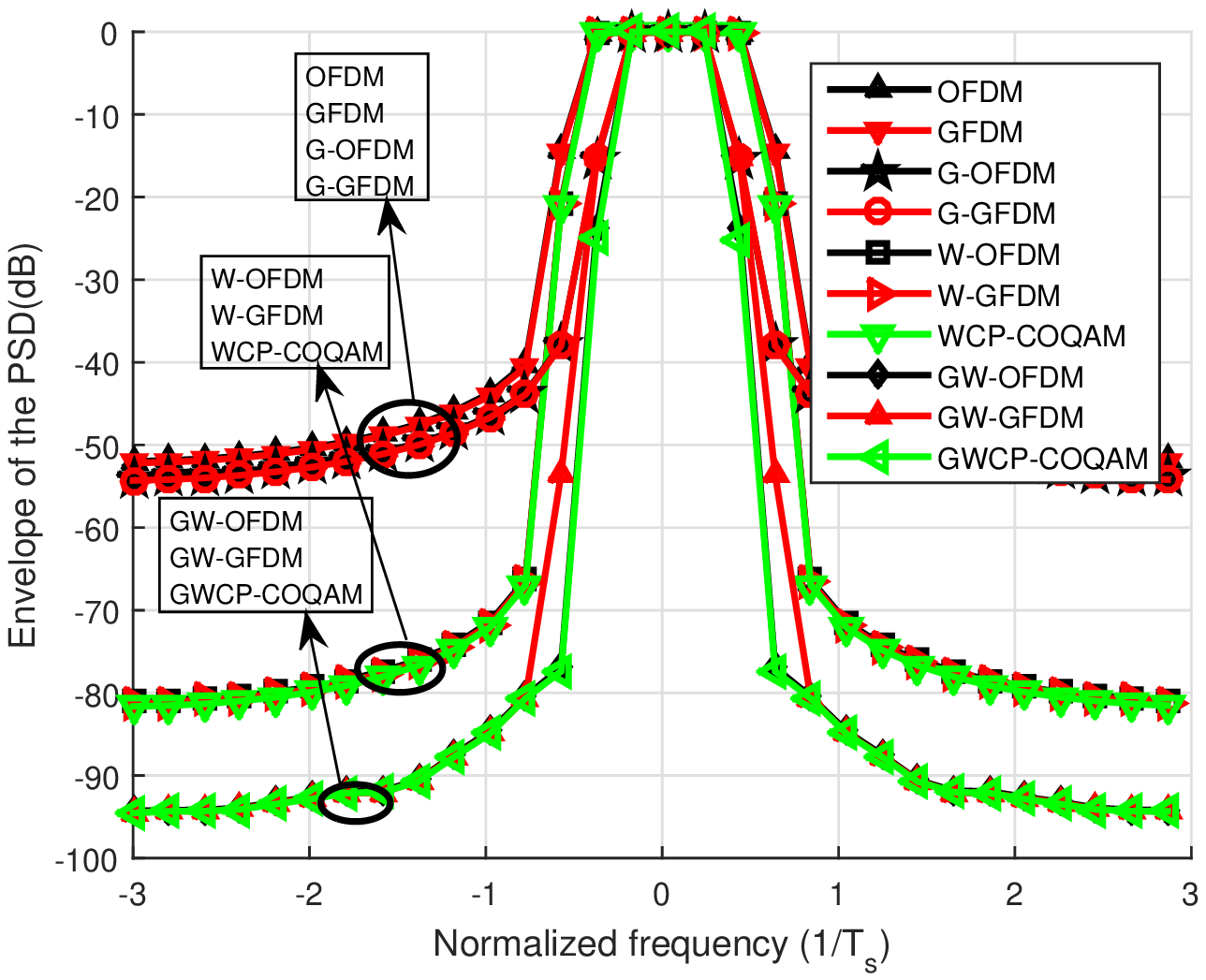}
\caption{\footnotesize PSDs for equal spectral efficiency when CP length=64 samples.}
\label{fig:OOB_1024OFDM}
\end{minipage}
\end{figure}
\underline{Error Rate Performance under CFO:}

\begin{figure}[H]
\centering
\graphicspath{ {C:/Bulut/Yuksek_Lisans/Yayinlar/OOB_GFDM/CFO_plots_comm_lett/CFO_plots_v2/} }
\includegraphics[width=0.6\columnwidth]{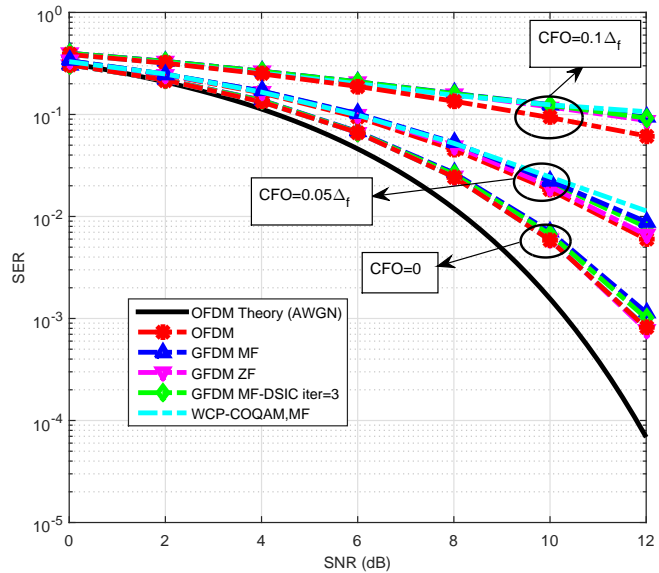}
\caption{Symbol error rate (SER) vs. SNR under CFO for OFDM, GFDM and WCP-COQAM when CP length=64 samples.}
\label{fig:OOB_1024OFDM}
\end{figure}

\pagebreak
\subsubsection{\textbf{Oversampling Rate=4}}
~\\ \hspace{30pt}\underline{OOB Emissions:}

\begin{figure}[H]
\centering
\graphicspath{ {C:/Bulut/Yuksek_Lisans/Yayinlar/OOB_GFDM/OOB_plots/OOB_plots_v2/} }
\includegraphics[width=0.6\columnwidth]{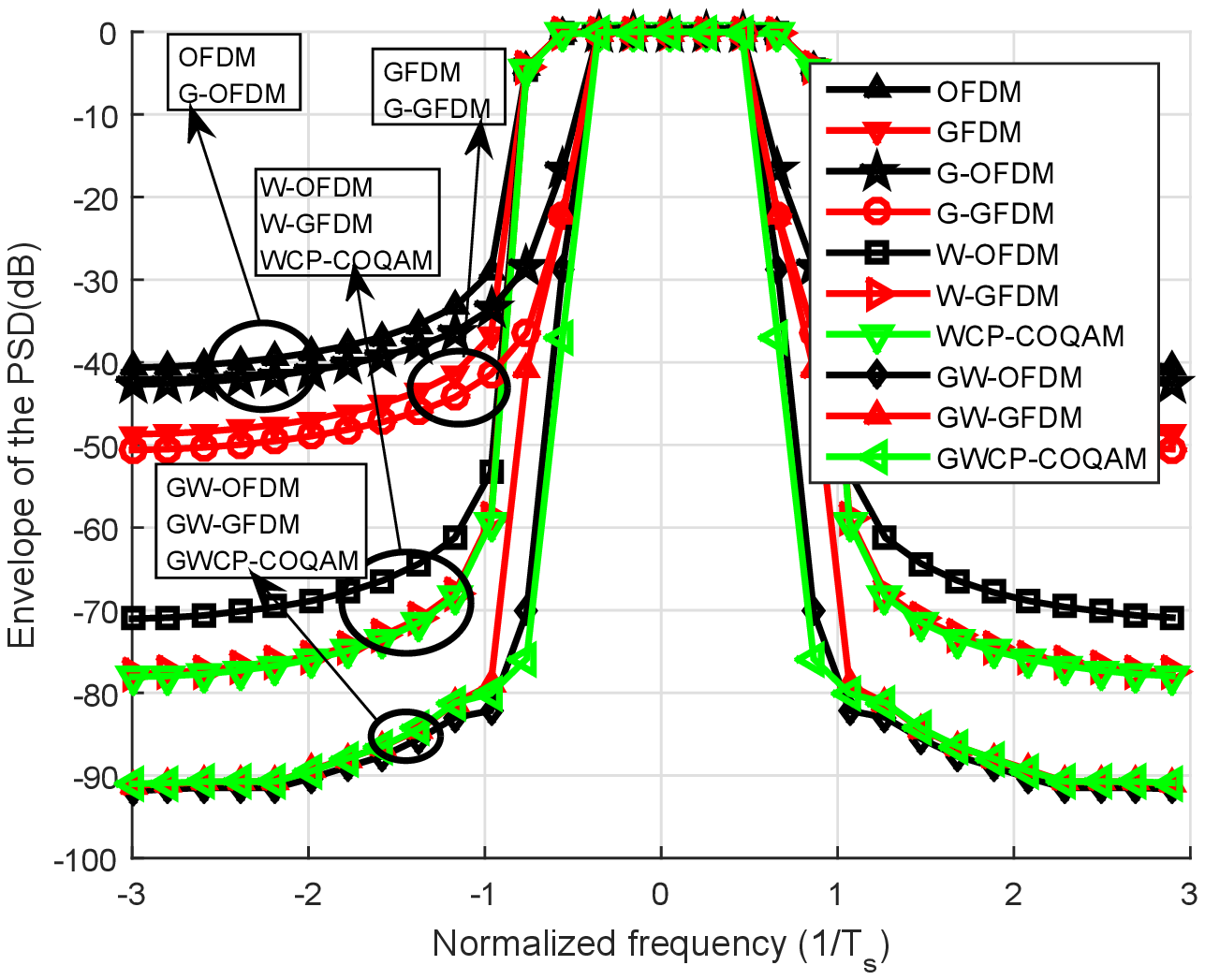}
\caption{PSDs for unequal spectral efficiency, non-contiguous case}
\label{fig:noncontig_uneq}
\end{figure}

\begin{figure}[H]
\centering
\graphicspath{ {C:/Bulut/Yuksek_Lisans/Yayinlar/OOB_GFDM/OOB_plots/OOB_plots_v2/} }
\includegraphics[width=0.6\columnwidth]{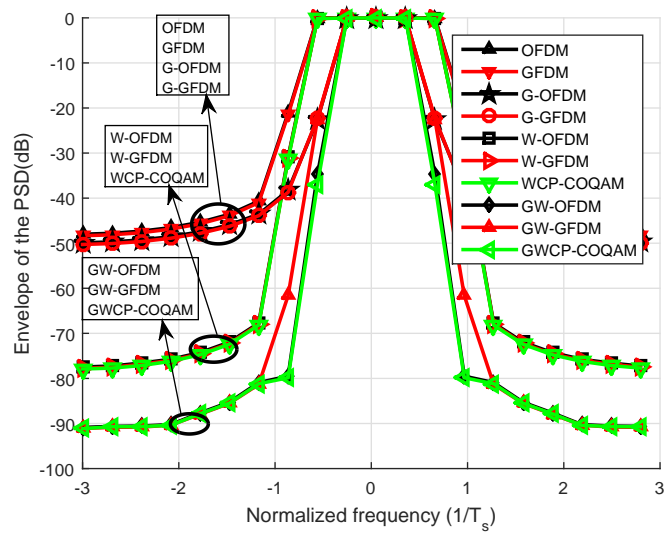}
\caption{PSDs for equal spectral efficiency, non-contiguous case}
\label{fig:contig_uneq}
\end{figure}

Here no CFO error rate curve is included since oversampling only affects the OOB spectrum. No oversampling is performed in CFO error rate simulations.

\pagebreak
\subsubsection{\textbf{Oversampling Rate=10}}
~\\ \hspace{30pt}\underline{OOB Emissions:}

\begin{figure}[H]
\centering
\graphicspath{ {C:/Bulut/Yuksek_Lisans/Yayinlar/OOB_GFDM/OOB_plots/OOB_plots_v2/} }
\includegraphics[width=0.6\columnwidth]{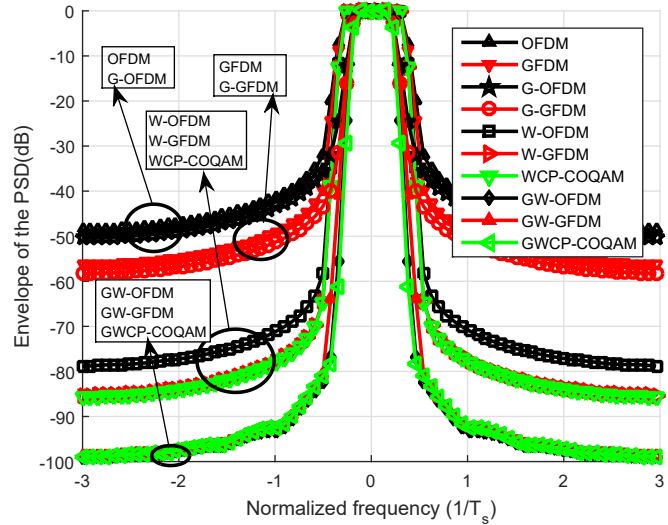}
\caption{PSDs for unequal spectral efficiency, non-contiguous case}
\label{fig:noncontig_uneq}
\end{figure}

\begin{figure}[H]
\centering
\graphicspath{ {C:/Bulut/Yuksek_Lisans/Yayinlar/OOB_GFDM/OOB_plots/OOB_plots_v2/} }
\includegraphics[width=0.6\columnwidth]{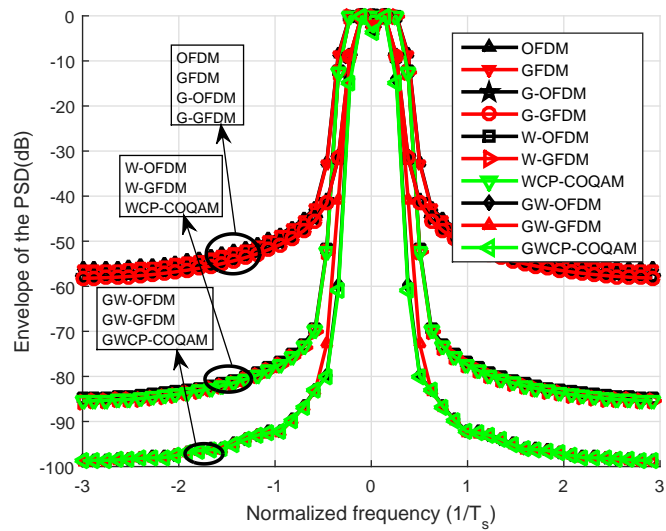}
\caption{PSDs for equal spectral efficiency, non-contiguous case}
\label{fig:contig_uneq}
\end{figure}

Here no CFO error rate curve is included since oversampling is only used to see the OOB spectrum. No oversampling is performed in CFO error rate simulations.

\pagebreak
\subsubsection{\textbf{Roll-off factor=0.4 (for the RC pulse shaping filter for GFDM and WCP-COQAM)}}
~\\ \hspace{30pt}\underline{OOB Emissions:}
\begin{figure}[H]
\centering
\begin{minipage}[b]{0.49\linewidth}
\graphicspath{ {C:/Bulut/Yuksek_Lisans/Yayinlar/OOB_GFDM/OOB_plots/OOB_plots_v2/} }
\includegraphics[width=\columnwidth]{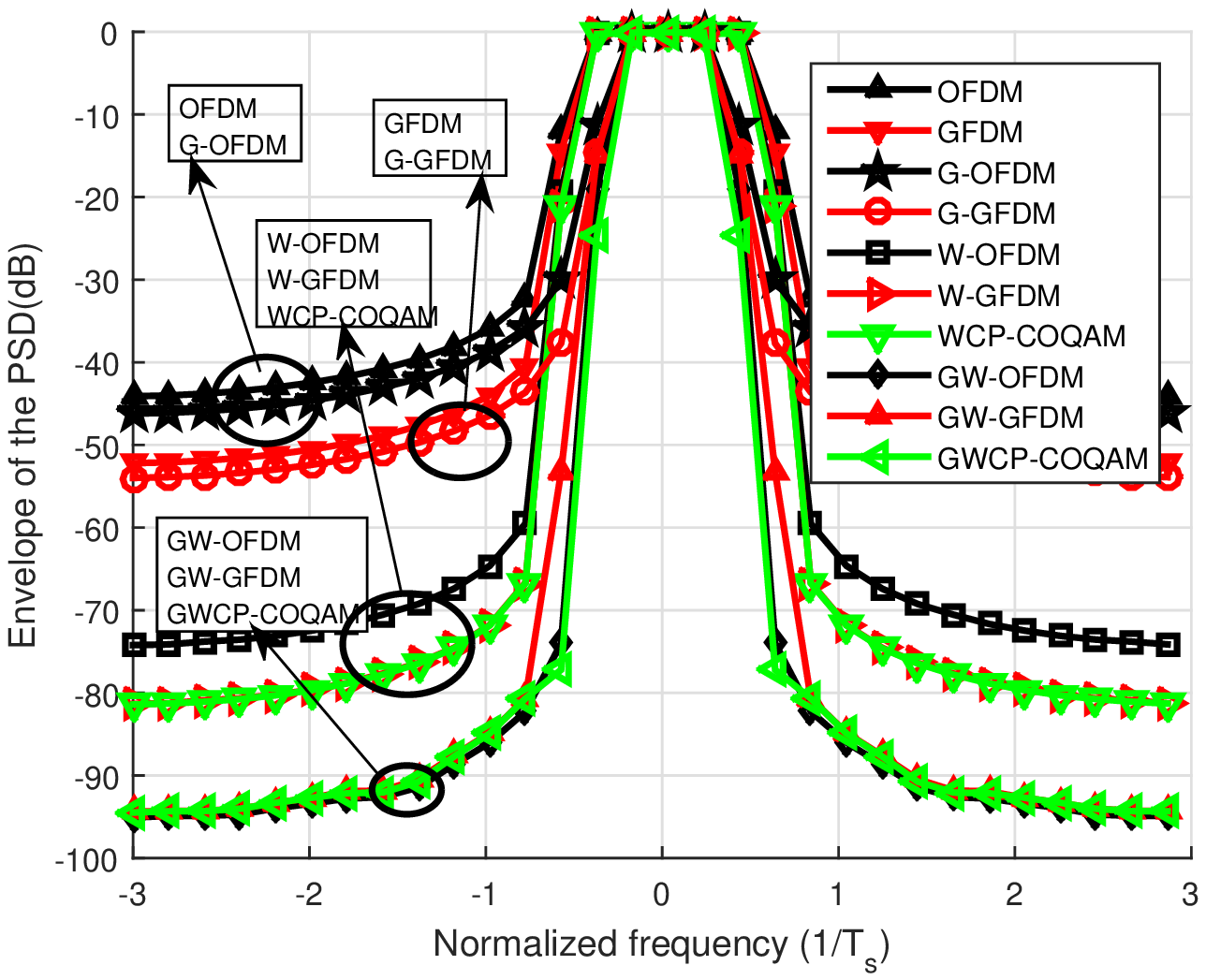}
\caption{\footnotesize PSDs for unequal spectral efficiency when roll-off factor for RC filter=0.4 .}
\label{fig:OOB_1024OFDM}
\end{minipage}
\begin{minipage}[b]{0.49\linewidth}
\graphicspath{ {C:/Bulut/Yuksek_Lisans/Yayinlar/OOB_GFDM/OOB_plots/OOB_plots_v2/} }
\includegraphics[width=\columnwidth]{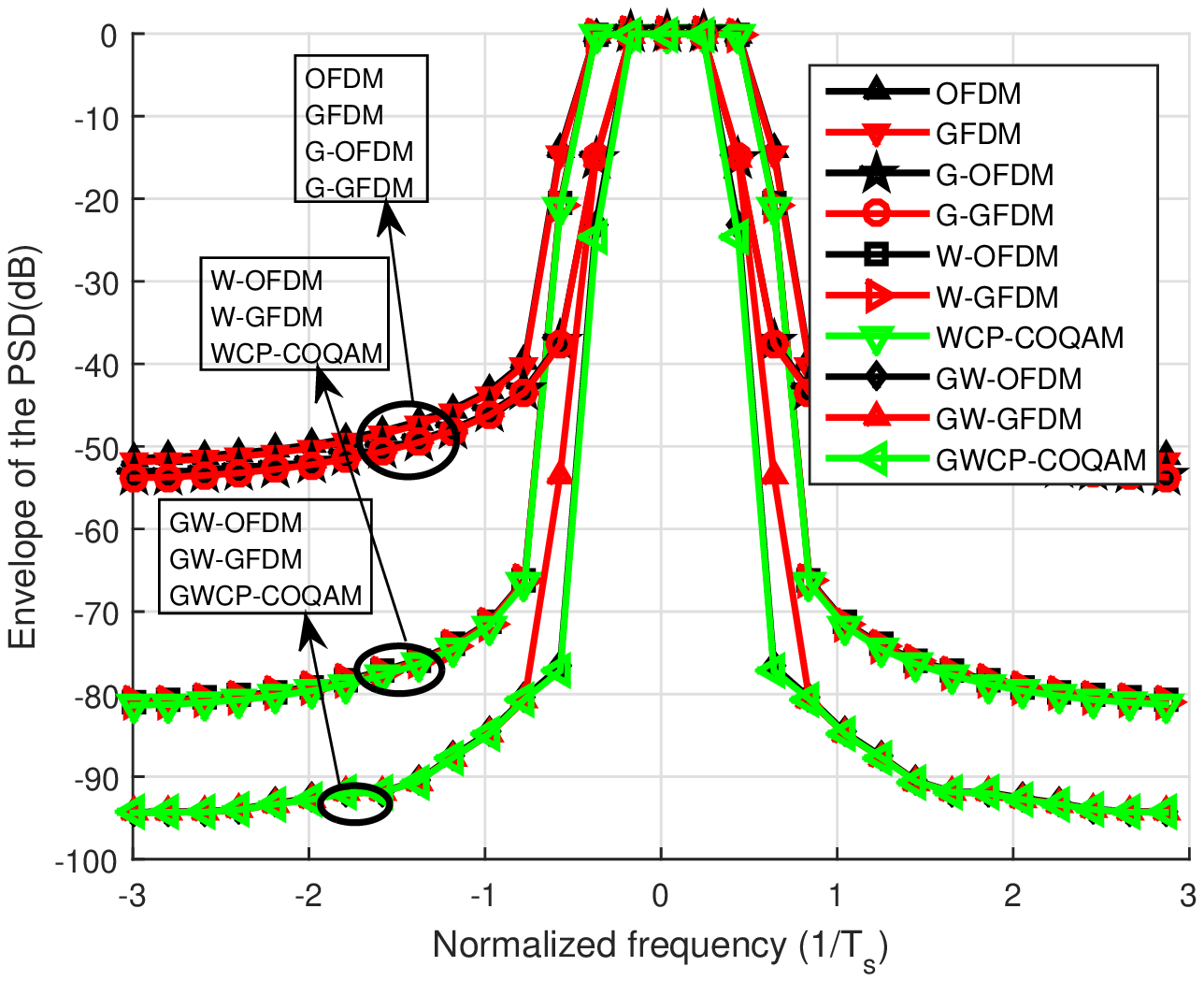}
\caption{\footnotesize PSDs for equal spectral efficiency when roll-off factor for RC filter=0.4 .}
\label{fig:OOB_1024OFDM}
\end{minipage}
\end{figure}
\underline{Error Rate Performance under CFO:}

\begin{figure}[H]
\centering
\graphicspath{ {C:/Bulut/Yuksek_Lisans/Yayinlar/OOB_GFDM/CFO_plots_comm_lett/CFO_plots_v2/} }
\includegraphics[width=0.6\columnwidth]{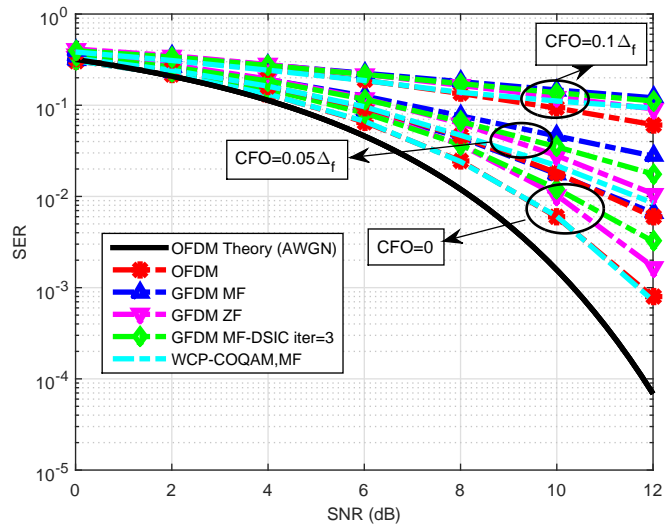}
\caption{Symbol error rate (SER) vs. SNR under CFO for OFDM, GFDM and WCP-COQAM when roll-off factor for RC filter=0.4.}
\label{fig:OOB_1024OFDM}
\end{figure}

\pagebreak

\subsubsection{\textbf{Constellation size=16 QAM}}
~\\ \underline{OOB Emissions:}
\begin{figure}[H]
\centering
\begin{minipage}[b]{0.49\linewidth}
\graphicspath{ {C:/Bulut/Yuksek_Lisans/Yayinlar/OOB_GFDM/OOB_plots/OOB_plots_v2/} }
\includegraphics[width=\columnwidth]{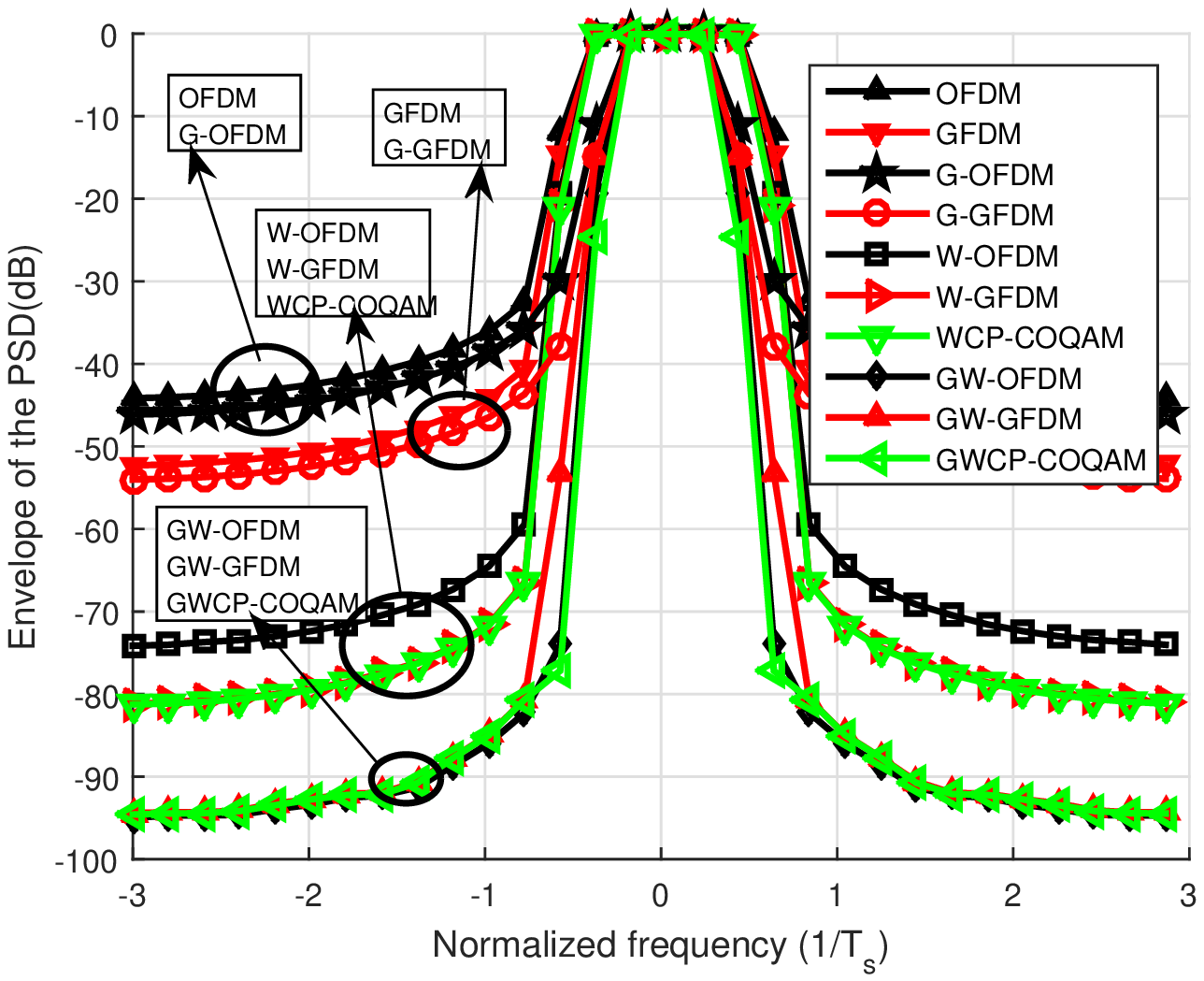}
\caption{\footnotesize PSDs for unequal spectral efficiency when constellation is 16-QAM.}
\label{fig:OOB_1024OFDM}
\end{minipage}
\begin{minipage}[b]{0.49\linewidth}
\graphicspath{ {C:/Bulut/Yuksek_Lisans/Yayinlar/OOB_GFDM/OOB_plots/OOB_plots_v2/} }
\includegraphics[width=\columnwidth]{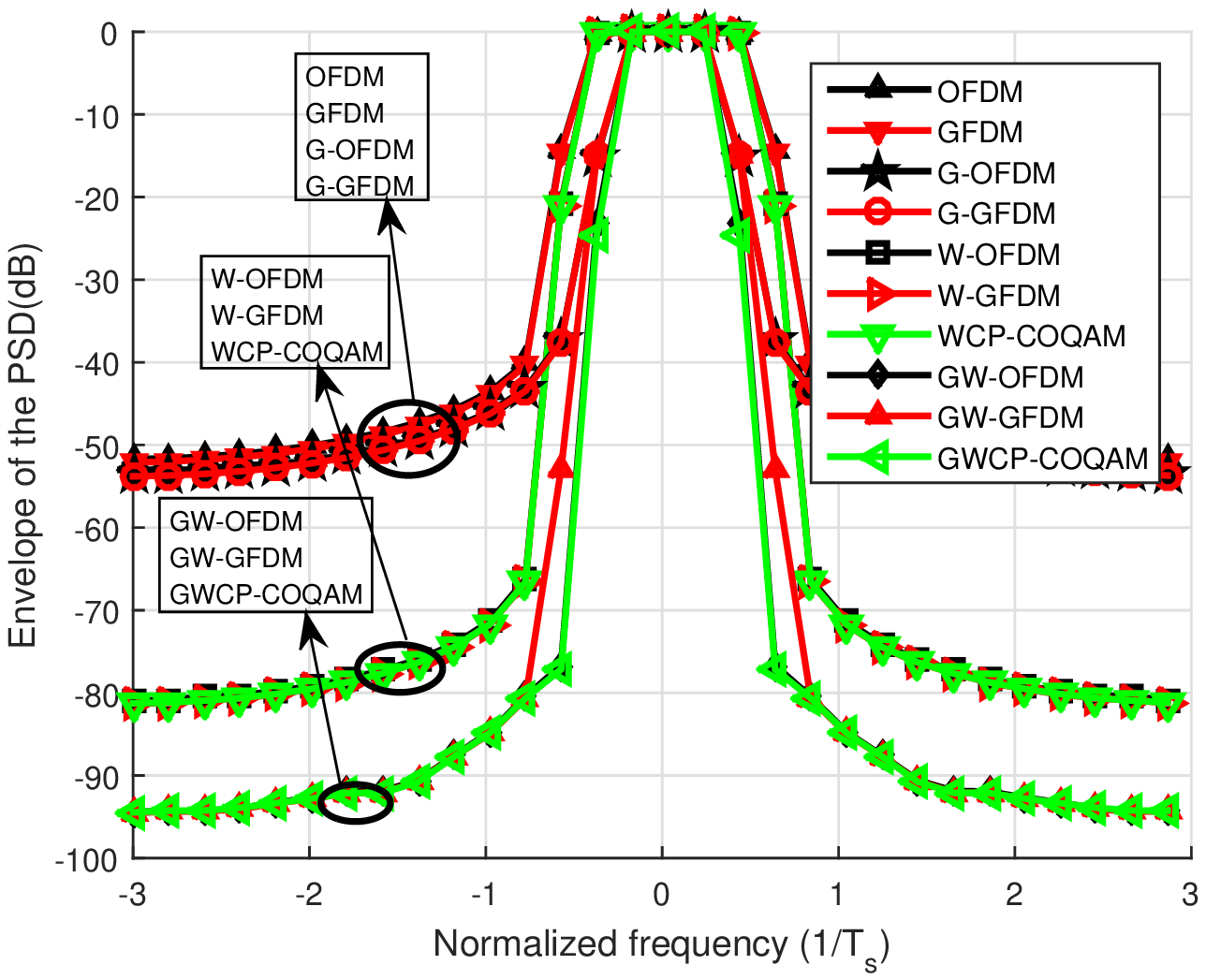}
\caption{\footnotesize PSDs for equal spectral efficiency when constellation is 16-QAM.}
\label{fig:OOB_1024OFDM}
\end{minipage}
\end{figure}
\underline{Error Rate Performance under CFO:}

\begin{figure}[H]
\centering
\graphicspath{ {C:/Bulut/Yuksek_Lisans/Yayinlar/OOB_GFDM/CFO_plots_comm_lett/CFO_plots_v2/} }
\includegraphics[width=0.6\columnwidth]{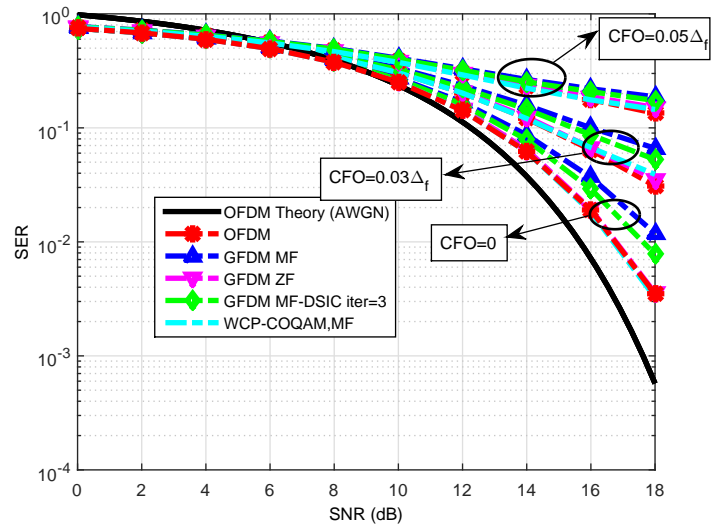}
\caption{Symbol error rate (SER) vs. SNR under CFO for OFDM, GFDM and WCP-COQAM when constellation is 16-QAM.}
\label{fig:OOB_1024OFDM}
\end{figure}
\pagebreak

\section{Non-contiguous Spectrum Case}
In this section, 2 non-contiguous channels are used. For each channel, the transmission parameters are as in Table~\ref{table:sim_param}. The bandwidth of the spacing between the non-contiguous channels are the same as the bandwidth of one of the non-contiguous channels.

\begin{figure}[H]
\centering
\graphicspath{ {C:/Bulut/Yuksek_Lisans/Yayinlar/OOB_GFDM/OOB_plots/OOB_plots_v2/} }
\includegraphics[width=0.55\columnwidth]{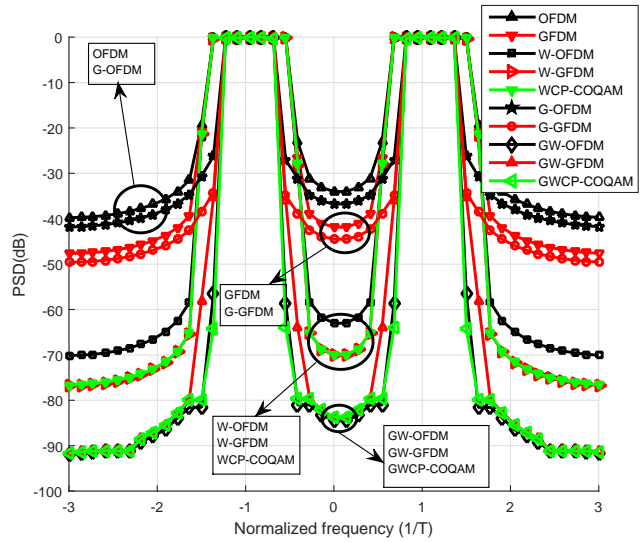}
\caption{PSDs for unequal spectral efficiency, non-contiguous case}
\label{fig:noncontig_uneq}
\end{figure}

\begin{figure}[H]
\centering
\graphicspath{ {C:/Bulut/Yuksek_Lisans/Yayinlar/OOB_GFDM/OOB_plots/OOB_plots_v2/} }
\includegraphics[width=0.55\columnwidth]{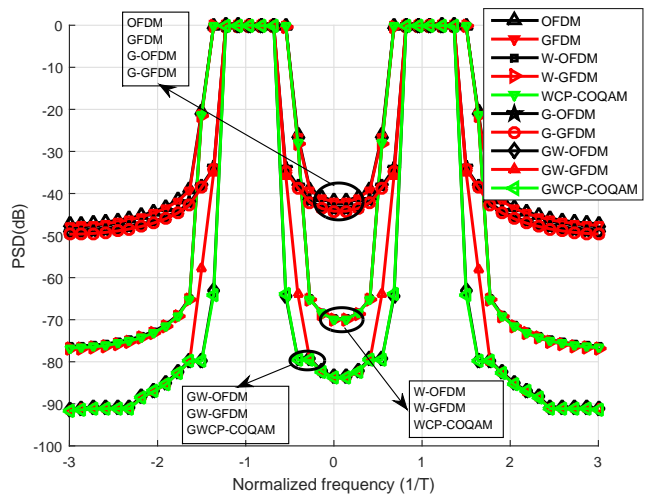}
\caption{PSDs for equal spectral efficiency, non-contiguous case}
\label{fig:contig_uneq}
\end{figure}

\section{Conclusion}
In this paper, GFDM, OFDM and WCP-COQAM are compared in terms of OOBE emissions for a wide range of simulation scenarios. The most important conclusion is that OFDM, GFDM and WPC-COQAM have similar OOBEs under equal spectral efficiency conditions. The fairness in the OOBE comparisons under equal spectral efficiency is also verified with the CFO immunity comparisons.

%
%

\ifCLASSOPTIONcaptionsoff
  \newpage
\fi

\bibliographystyle{IEEEtran}
\bibliography{IEEEabrv,referans_aux}

\end{document}